\documentclass[12pt]{article}
\usepackage{epsfig}
\newcommand{\bmat}{\left(\begin{array}}
\newcommand{\emat}{\end{array}\right)}
\def\NPB#1#2#3{Nucl. Phys. B{#1} (19#2) #3}
\def\PLB#1#2#3{Phys. Lett. B{#1} (19#2) #3}

\def\PRD#1#2#3{Phys. Rev. D{#1} (19#2) #3}

\def\yzero{\smash{\hbox{$y\kern-4pt\raise1pt\hbox{${}^\circ$}$}}}

\def\-{\hphantom{-}}

\def\s2{\frac{1}{\sqrt2}}

\def\beq{\begin{equation}}
\def\eeq{\end{equation}}
\def\beqa{\begin{eqnarray}}
\def\eeqa{\end{eqnarray}}

\def\Tr{{\rm Tr \,}}

\def\IF{\relax{\rm I\kern-.18em F}}
\def\II{\relax{\rm I\kern-.18em I}}
\def\IP{\relax{\rm I\kern-.18em P}}
\def\IC{\relax\hbox{\kern.25em$\inbar\kern-.3em{\rm C}$}}
\def\IR{\relax{\rm I\kern-.18em R}}

\def\Dsl{\,\raise.15ex\hbox{/}\mkern-13.5mu D} 
\def\IZ{Z\kern-.4em  Z}
\def\bmat{\left(\begin{array}}
\def\emat{\end{array}\right)}
\def\id{{\rm I}}
\def    \part          {\partial}
\def    \be            {\begin{equation}}
\def    \ee            {\end{equation}}
\def    \bea           {\begin{eqnarray}}
\def    \eea           {\end{eqnarray}}
\def    \nn            {\nonumber}

\def    \ccii          {C_i^{5_i}}

\def    \ccia          {C_1^{5_i}}  
\def    \ccib          {C_2^{5_i}}  
\def    \ccic          {C_3^{5_i}}  
\def    \ccij          {C_{j}^{5_{i}}}

\def    \ccjk          {C_{j}^{5_{k}}}

\def    \cni           {C_i^{9}}  
\def    \cna           {C_1^{9}}
\def    \cnb           {C_2^{9}} 
\def    \cnc           {C_3^{9}}
\def    \cnci          {C^{9 5_i}}
\def    \cncj          {C^{9 5_{j}}}   
\def    \cnck          {C^{9 5_{k}}}

\def    \ccjck         {C^{5_{j} 5_{k}}}
\def    \ccick         {C^{5_{i} 5_{k}}}
\def    \ccicj         {C^{5_{i} 5_{j}}}
\def    \ccacb         {C^{5_1 5_2}}
\def    \cccca         {C^{5_3 5_1}}
\def    \ccbcc         {C^{5_2 5_3}}
\begin{document}
%
\makeatletter
\@addtoreset{equation}{section}
\makeatother
\renewcommand{\theequation}{\thesection.\arabic{equation}}
\pagestyle{empty}
\rightline{FTUAM 98/28}
\rightline{IFT-UAM/CSIC-98-28}
\vspace{0.5cm}
\begin{center}
\LARGE{Aspects of
Type I String Phenomenology \\[20mm]}
\large{L.E. Ib\'a\~nez, C. Mu\~noz and S. Rigolin\\[8mm]}
\small{
Departamento de F\'{\i}sica Te\'orica C-XI
and Instituto de F\'{\i}sica Te\'orica  C-XVI,\\[-0.3em]
Universidad Aut\'onoma de Madrid, 
Cantoblanco, 28049 Madrid, Spain. 
\\[7mm]}
\small{\bf Abstract} 
\\[7mm]
\end{center}
\begin{center}
\begin{minipage}[h]{14.0cm}
We study different phenomenological aspects of compact, $D=4$, $N=1$ 
Type IIB  orientifolds considered  as models  for unification of the 
standard model and gravity. We discuss the structure of the compactification,
string and unification scales depending on the different possible D-brane 
configurations. It is emphasized that in the context of Type I models
the $M_W/M_{Planck}$ hierarchy problem is substantially alleviated
and may be generated by geometrical factors. We obtain the  effective 
low-energy supergravity Lagrangian and derive the form of soft SUSY-breaking 
terms under the assumption of dilaton/moduli dominance. 
We also discuss the role of anomalous $U(1)$'s and of twisted moduli
in this class of theories. A novel mechanism based on the role of 
singularities is suggested to achieve consistency with gauge coupling 
unification in low  string scale models.
\end{minipage}
\end{center}
\newpage
\setcounter{page}{1}
\pagestyle{plain}
\renewcommand{\thefootnote}{\arabic{footnote}}
\setcounter{footnote}{0}
%
%
\section{Introduction}
In the attempts to embed the known standard model (SM) interactions
within perturbative $D=4$, $N=1$ heterotic vacua, a number of
properties were  generally assumed \cite{rev}.
The string scale $M_{string}$ was identified essentially with the Planck 
mass $M_{Planck}$ and the gauge coupling unification scale $M_X$ was also 
taken to be in the vicinity of $M_{string}$. These scale identifications 
were not arbitrary, they were essentially dictated by the structure of 
the heterotic string couplings. The SM gauge group was a subgroup of
either $E_8\times E_8$ or $SO(32)$ and SUSY breaking was most
of the times assumed to take place  by gaugino condensation in
a hidden sector of the theory. A number of interesting 
four-dimensional $N=1$ vacua with particle content not far from that of 
the supersymmetric SM were found \cite{rev}.

With the string theory developments of the last three years it
has been realized that many of the assumptions for embedding the
SM inside string theory were in fact mere artifacts of {\it perturbative} 
heterotic string compactifications. New  vacua based on Type II and Type I 
strings have revealed the importance of Dirichlet D-branes \cite{polrev}
in the understanding of the general space of string vacua. In addition all 
the five known perturbative SUSY strings seem to correspond (along with 
$D=11$ supergravity) to particular limits of the unique underlying M-theory.

We thought we knew what the size of the string scale was: just close to 
$M_{Planck}$. We have now realized that we didn't. 
Going away from perturbative heterotic vacua $M_{string}$ is related both 
to $M_{Planck}$ and the compactification scale $M_c$ in such a way that, 
in principle, $M_{string}$ may be anywhere between the weak scale and the 
Planck scale without any obvious contradiction with experimental facts 
\cite{witten}--\cite{bursta}. 
In addition there is a much larger ambiguity in the possibilities for gauge 
group for $D=4$, $N=1$ vacua. Gauge groups live on the world-volume of 
Dp-branes and the observed SM gauge group may in principle correspond to, 
e.g., gauge interactions on the world-volume of 3-branes (or D(n+3)-branes 
with n dimensions wrapping on some compact space).

Although we are far from a complete understanding of the general structure 
of the new $D=4$, $N=1$ string vacua now available, it is perhaps time 
to try to extract some general characteristics from the known examples
and see how the traditional perturbative heterotic schemes are 
changed. We will be interested here in {\it compact}  $D=4$, $N=1$ string 
vacua. We are aware essentially of three avenues to construct such new 
vacua: 1) Type IIB $D=4$ compact orientifold models, 2) M-theory 
compactification on CY$\times S^1/Z_2$  and 3) F-theory compactifications 
on CY four-folds. 

The first is the natural generalization of the toroidal heterotic 
orbifolds studied in the past \cite{dhvw,orbi} to the Type IIB and Type I 
string theories. Their construction has developed in the last two years
\cite{bl}--\cite{lpt} and provides us with new perturbative vacua. 
Due to Type I/heterotic duality, some of these 
new vacua may be understood as dual to other {\it  non-perturbative } 
heterotic vacua. This is the class of theories we are going to concentrate 
on in the present article. We think they are interesting because they are 
perturbative vacua, with which one can use standard string theory techniques,
and yet provide interesting $D=4$, $N=1$ models with chiral matter. 
Furthermore, due to Type I/heterotic duality one expects to extract from 
them information about the strongly coupled regime of heterotic vacua.

We should emphasize that the model building in this kind of Type IIB 
orientifolds is still in its infancy and there are not yet fully 
realistic models. Our hope is rather trying to extract some {\it generic 
features} of this kind of vacua hoping that could be shared by more 
realistic models based on these (or other D-brane) constructions.

The content of the article is as follows. In section 2 we present 
a general overview of the structure of Type IIB  compact $D=4$ 
orientifold models and of the important role played by D-branes and 
T-dualities in their construction. In section 3 we study the general 
structure of mass scales in this class of theories. After reviewing the 
relationship between string scale, Planck mass and compactification scale,
we analyze, in subsection 3.2, different possibilities to embed the SM 
interactions within some Dp-brane sector with unification of coupling 
constants at $M_X=2\times 10^{16}$ GeV, as suggested by plain
extrapolation of low-energy data. We find that, in order to do that within 
an isotropic compactification, one has to identify the unification scale 
$M_X$ with the string scale for 3-branes or 9-branes but one has to 
identify it with a compactification/winding scale for 5/7 branes. We also
consider the possibility of embedding the SM $SU(3)_C$ and $SU(2)_L$ 
interactions within different types of Dp-branes. In section 4 we discuss 
the generic presence of very weakly coupled gauge and Yukawa interactions 
in models with the string scale below the Planck mass and their possible 
phenomenological applications.

In section 5 we discuss possible avenues for the generation of
the notorious $M_W/M_{Planck}$ hierarchy. We argue that in the context 
of Type I models this problem is substantially weaker and the hierarchy may 
appear from purely geometrical factors.

In section 6  we extract the general form of the tree-level K\"ahler 
potential and gauge kinetic functions for this class of Type IIB 
orientifolds. In order to do that we make use of the T-duality 
symmetries which exchange among themselves the different types of
Dp-branes. These symmetries also exchange dilaton with moduli so that 
all are essentially on equal footing.
As a consequence the K\"ahler potential is rather symmetric under the
exchange of dilaton and moduli. Furthermore the gauge kinetic function
is also linear in either the dilaton $S$ or moduli $T_i$, depending on
what Dp-brane the gauge group is originated from. All this is drastically 
different from the perturbative heterotic vacua in which the complex dilaton 
$S$ has a unique universal role.

In section 7 we use the results of section 6 in order to 
derive general patterns of soft SUSY-breaking terms under the 
assumption of dilaton/moduli SUSY breaking \cite{BIM22}--\cite{BIMS}. 
We emphasize that this assumption of 
dilaton/moduli dominance is more compelling in the D-brane scenarios 
where only closed string fields like $S$ and $T_i$ can move into 
the bulk and transmit SUSY breaking from one D-brane sector to some other.
The soft terms obtained present explicit invariances under T-duality
transformations. In certain schemes dilaton dominance is thus T-dual 
(and hence equivalent) to modulus dominance. As happened for similar soft 
terms in untwisted sectors of heterotic orbifolds, certain sum-rules among 
soft terms are fulfilled for any goldstino direction.

In section 8 we  describe some specific issues regarding  the role of 
pseudo-anomalous $U(1)$'s within the context of Type IIB $D=4$, $N=1$ 
orientifolds. In this class of theories the gauge kinetic functions have 
also explicit dependence on the twisted moduli fields associated to the 
orbifold singularities \cite{iru}. We find that in certain class of 
orientifolds the coefficient of that dependence is proportional to the 
beta function of the group considered. This suggests a new and natural 
way to achieve gauge coupling unification in models with low (or very low)
string scale. We leave section 9 for some general final comments. We point 
out that, if a generic vacuum string configuration has $N=0$ brane sectors 
(from e.g., D-branes wrapping on non-supersymmetric cycles in the compact 
dimensions), will be necessary to reduce the string scale well below the 
Planck mass for avoiding large SUSY-breaking contributions in the ($N=1$) 
brane sector containing the SM.

Previous studies concerning possible phenomenological aspects of Type IIB
compact orientifolds may be found in refs.\cite{lykken,
ibanez,afiv,2kakus,shiutye,kakutye,benakli,BIQ,kakutev}. 
%
%
\section{Type IIB orientifolds, Dp-branes and T-duality}
Ten-dimensional $SO(32)$ Type I string theory may be understood as
an ``orientifold'' \cite{sagnotti}--\cite{gj} of $D=10$, Type IIB theory 
under the $Z_2$ operation designated world-sheet parity $\Omega $
\cite{dab}. Let $(\sigma,\tau )$ be the two (space-like and time-like)
world-sheet coordinates of the string. Defining the complex
world-sheet coordinate $z=\exp(\tau + i\sigma )$, one then has
$\Omega z={\overline z}$. Thus $\Omega $ transforms left-moving and 
right-moving vibrations of the string into each other, so that the 
result of a projection of IIB string under $\Omega $ is a closed 
unoriented string with only one, $D=10$, SUSY. In addition it turns 
out that the consistency of the theory (tadpole cancellation)
requires the addition of twisted sectors with respect to
$\Omega $. These are nothing else but Type I open strings which have
to be added to the closed unoriented strings discussed above.
Type IIB strings possess extended solitonic objects, Dp-branes,
which have  Ramond-Ramond (R-R) charges. For our purposes
they may be defined as submanifolds of the full $D=10$ space
where the open strings can end. They have world-volumes spanning $p+1$ 
dimensions with $p=-1,1,3,5,7,9$ \cite{polrev} . Here we will only 
be concerned with the cases $p=3,5,7,9$. Open strings can only 
start or end on p-branes. 
An open string has Dirichlet boundary conditions \cite{polrev} in the 
$9-p$ coordinates transverse to the p-brane whereas it has Neumann 
boundary conditions in the world-volume directions. In $D=10$ the overall 
cancellation of the 
R-R charge (equivalent to tadpole cancellation) requires the presence of 
32 9-branes in the vacuum. Open strings ending on 9-branes (which means 
in this case anywhere) will give rise to massless $D=10$ gauge fields in 
the adjoint of $SO(32)$.  

Type IIB, $D=4$, $N=1$ toroidal orientifolds \cite{bl}--\cite{lpt} are 
constructed by compactifying Type IIB theory on a six-torus $T^6$ and
further moding the theory under some discrete symmetry $G$ which acts 
as a discrete rotation on the compact $SO(6)$ spacetime symmetry. Thus we 
will have an orientifold of Type IIB on $T^6$ moded by $\{\Omega \times G\}$. 
We will consider Abelian modings with $G=Z_N$ or $Z_N\times Z_M$ in such a 
way that there is $N=1$ SUSY in four dimensions. We refer the reader to e.g. 
ref.\cite{afiv} for a classification of possible discrete groups and 
further technical details. This procedure leads to overall non-vanishing 
R-R charges. In order to obtain consistent vacua we have to
add certain types and numbers of p-branes in the vacuum
in such a way that the overall charge vanishes. The type
and number of p-branes required will depend on the particular 
orientifold considered. For $Z_N$ modings with odd $N$, it turns out
that only 9-branes may be present. That is the case of the 
$Z_3,Z_7$ and $Z_3\times Z_3$ orientifolds, the only ones allowed 
by D=4, N=1 SUSY. For even $N$, one can have both $9$-branes 
and/or $5$-branes. The 5-branes will have their world-volume spanning
Minkowski space plus one complex compact dimension $X_i$, i=1,2,3.
Thus there may be up to three types of 5-branes which will be
denoted by  $5_i$, i=1,2,3 depending on what complex compact coordinate 
is included in the 5-brane world-volume.

Instead of $\Omega $ one can use other $Z_2$ modings which are still 
consistent with  $N=1$ SUSY in $D=4$. Let $O_i$ denote a reflection of 
the $i-th$ compact complex coordinate, i.e., $O_i(X_j)=-X_i$
for $i=j$, $+X_j$ for $i\not= j$.
In the cases discussed above one has as orientifold generators
$\Omega $ and (for even $N$, in which case $G$ contains an $O_iO_j$, 
$i\not = j$ twist) $\Omega O_iO_j$. One can also construct consistent 
$D=4$, $N=1$ orientifolds using instead $(-1)^{F_L}\Omega O_i $ and 
$(-1)^{F_L}\Omega O_iO_jO_k$, $i\not = j\not = k\not = i$.
Here $F_L$ is the world-sheet left-handed fermion number.
In this case it turns out that tadpole cancellation conditions
will require in general the presence in the vacuum of 7-branes and
3-branes respectively. There may be three different types of 
7-branes, $7_i$, $i=1,2,3$ depending what complex dimension $X_i$ is
transverse to the 7-brane world-volume.

Thus we see that, depending on the orientifold generators, one
can deal with 3-branes, $5_i$-branes, $7_i$-branes 
and 9-branes. Not all types may be present simultaneously if 
we want to preserve $N=1$ in $D=4$. For a given $D=4$, $N=1$ 
vacuum with D-p-branes and D-p$'$-branes one must have
$(p-p')=0,\pm 4$. Thus we can have at most either 9-branes 
with $5_i$-branes {\it or} 3-branes with $7_i$-branes.
The number of each type of p-brane in each case is dictated
by tadpole cancellation constraints. These in turn guarantee
the cancellation of gauge anomalies in the effective 
$D=4$, $N=1$ theory. Notice that there will be a gauge group
associated to each set of coincident p-branes of a given type.

T-dualities relate the different types of p-branes present in each given 
vacuum \cite{polrev}. Consider for simplicity the 6-torus as the product of
three two-tori, $T^6=T^2\times T^2\times T^2$ each with compact radii
$R_i$, $i=1,2,3$. Now, it is well known that
a duality transformation $R_i\rightarrow \alpha '/R_i$ transforms
Neumann boundary conditions on the $X_i$ coordinate into Dirichlet
boundary conditions and vice versa \cite{polrev}. This means that e.g., 
a 9-brane will turn into a $7_i$-brane and vice versa under this 
transformation. More generally, consider a general configuration with 
different types of p-branes. T-dualities will have the effect:
\beqa
    R_1\leftrightarrow \alpha'/R_1 \ \ 
 & R_{1,2}\leftrightarrow \alpha '/R_{1,2}
 & \ \   R_{1,2,3} \leftrightarrow \alpha ' 
/R_{1,2,3}   \nonumber \\
    9 \leftrightarrow 7_1 \ \   & 9\leftrightarrow 5_3 &\ \  
9\leftrightarrow 3  \nonumber \\
7_2\leftrightarrow 5_3 \ \ & 7_{1,2}\leftrightarrow 7_{2,1} &
\ \ 7_1\leftrightarrow 5_1 \nonumber \\
7_3\leftrightarrow 5_2 \ \ & 7_3\leftrightarrow 3 &    
\ \ 7_2\leftrightarrow 5_2 \nonumber \\
5_1\leftrightarrow 3 \ \ & 5_{1,2}\leftrightarrow 5_{2,1} &    
\ \ 7_3\leftrightarrow 5_3 \ . 
\label{pduales}
\eeqa
In addition one has to rescale the Type I coupling $\lambda _I$
as:
\beq
\lambda _I\leftrightarrow \lambda _I({{\alpha '}\over {R_1^2} })\ ;\
\lambda _I\leftrightarrow \lambda _I({ {{\alpha '}^2} \over {R_1^2R_2^2} })\ ;\
\lambda _I\leftrightarrow \lambda _I({ {{\alpha '}^3} \over {R_1^2R_2^2R_3^2} })
\label{rescaleo}
\eeq
respectively, where $\alpha '=1/M_I^2$ with $M_I$ the Type I string mass. 
Thus given any configuration with certain distribution 
of p-branes in the vacuum, there are a number of equivalent
configurations which are obtained from T-dualities as shown above.

Given a p-brane in a background with six compact dimensions, open strings
ending on that p-brane will only have
Kaluza-Klein (KK) states along the compact dimensions with 
Neumann boundary conditions.  On the contrary, it will have
winding states only in those compact directions with 
Dirichlet boundary conditions. This will mean for example that
open strings ending on 9-branes will only have KK modes but no winding
modes. On the contrary, open strings ending on 3-branes will have
no  KK modes but will have winding modes along all compact directions.
These winding modes will correspond to open strings starting at the 
3-brane, going around the torus and coming back to the 3-brane
\cite{polrev}. 
In the same way open strings ending on a $5_i$-brane will have
KK modes on the $X_i$ direction but winding modes in the other two complex
dimensions. The opposite happens with open strings on $7_i$-branes:
they will have windings in the $X_i$ direction and KK modes in the
other two. On the other hand, closed strings can have both
KK and winding modes in all compact dimensions. This turns out to be
important in order to figure out what is the scale related to 
gauge coupling unification, as we will see in the next section.

Given different sets of p-branes in a given $D=4$, $N=1$ orientifold,
there will be a gauge group and charged chiral fields associated to each 
set of coincident (in the transverse dimensions) p-branes. They correspond 
to zero modes of open strings starting and ending on the {\it same} set 
of p-branes. In addition there will be in general chiral fields (but no 
gauge group) corresponding to open strings stretching between different 
types of p-branes like e.g. strings stretching between 9-branes and 5-branes. 
The different types of chiral matter fields appearing will be reviewed in 
section 6 (see e.g. ref.\cite{afiv} for specific examples of orientifold 
models). In addition to  gauge groups and charged chiral fields coming 
from open strings there are also closed strings singlet chiral fields.
Among those there will be the complex dilaton $S$ and the untwisted 
moduli fields $T_i$. These are also reviewed in section 6.

Let us finally remark some points about Type I/heterotic duality. 
It is believed that the strongly coupled limit of $D=10$ Type I
string is the heterotic $SO(32)$ string \cite{witpol}. Thus one expects 
that the Type IIB orientifold vacua here discussed will have 
heterotic duals. However, these heterotic duals need not correspond to
known perturbative heterotic vacua. Indeed, the orientifolds here discussed
often have gauge groups with rank bigger than 20, which is impossible
for perturbative heterotic vacua. Rather, the present class of
models should correspond to some non-perturbative points in the
heterotic moduli space. 
%
%
\section{D-branes, string scale and gauge coupling unification in
Type I $D=4$ string vacua}
%
%
\subsection{D-branes, string scale and Planck mass}
We will start in this section by studying  the relationship
between string, Planck and compactification scales in Type I $D=4$ 
strings of the type described above. Let us consider the relevant piece 
of the bosonic action of the $D=10$, $N=1$ effective Lagrangian appearing
in Type I string theory:
\beq
S_{10} \ =\ - \int {{dx^{10}}\over {(2\pi )^7}}
 \sqrt{-g}\  (\  {{ M_I^8}\over {\lambda _I^2}} \ R \ 
+\  {{M_I^6}\over {\lambda _I}} \ {1\over 4} F^2_{(9)} \ +\ ...\  )\ ,
\label{d10}
\eeq
where $\lambda _I$ is the $D=10$ Type I dilaton, $M_I=1/\sqrt{\alpha'}$ 
is the Type I string scale and the subindex $(9)$ refers to the gauge group 
coming from the 32 9-branes whose world-volume fills the whole ten-dimensional 
space. We consider now the dimensional reduction down to four dimensions 
obtained by compactification on an orbifold with an underlying compact 
torus of the form $T^2\times T^2\times T^2 $.  The three 
tori are taken with volumes $(2\pi R_i)^2$, $i=1,2,3$ respectively. 
One obtains \cite{witten,lykken,anton,shiutye,lpt}
\beqa
S_{4} \ &  = & \ - \int {{ dx^{4}}\over {2\pi }}
 \sqrt{-g}\  (\  {{R_1^2R_2^2R_3^2 M_I^8}\over {\lambda_I^2}} \ R \ + \  
{{R_1^2R_2^2R_3^2M_I^6}\over {\lambda _I}} \ {1\over 4} F^2_{(9)} \nn \\   
&  + & \ \sum _{i\not= j\not=k\not= i} {{R_i^2 R_j^2 M_I^4}\over 
{\lambda_I}} \ {1\over 4} F^2_{(7_k)} \  +  \ \sum _{j=1}^3 
{{R_j^2M_I^2}\over {\lambda _I}} \ {1\over 4}  F^2_{(5_j)} \ +
{1\over {\lambda _I}}\ {1\over 4}  F_{(3)}^2  \ +\ ...\  )\ ,
\label{d04}
\eeqa
where we have displayed the kinetic terms for gauge bosons of the different 
groups which may come from the different p-branes, $p=9, 7_k, 5_j, 3$. 
As discussed above, not all the different p-brane sectors 
should be present in the vacuum if we want to respect $N=1$ SUSY. 
The particular radius dependence of each p-brane may be obtained by noting 
that the gauge coupling $g_p$ of the vector bosons on a given p-brane is 
related with the Type I dilaton and the string scale as \cite{polrev}
\beq
  g_p^2\ =\ {{\lambda _I}\over {(2\pi )^{(2-p)}}} \ M_I^{(3-p)} \ .
\label{gaugec}
\eeq
Only for 3-branes one has that $g_3$ is dimensionless, while for $p=5,7,9$ 
one gets additional radius dependence corresponding to the fact that p-branes
are wrapping on 2, 4 or 6 compact coordinates respectively. From the above 
equations one obtains for the gravitational coupling $G_N$
\beq
G_N \ =\ {1 \over {M_{Planck}^2}} \ =\  { {\lambda_I^2 M_1^2 M_2^2 M_3^2} 
       \over {8 M_I^8  }}
\label{planck}
\eeq
and for the gauge couplings $\alpha_p $ for the different 
p-branes\footnote{We will see in section 8 that in the presence of
repaired orbifold singularities there are extra contributions to
the gauge coupling constants proportional to the blowing-up
modes of the singularities. This may have important effects 
which we discuss in that section.}:
\beqa
 \alpha _9 \ & = &\   { { \lambda _I M_1^2M_2^2M_3^2 } \over
{2 M_I^6 } }     \ ; \  
\alpha _{7_i} \ =\    { { \lambda _I M_j^2M_k^2 } \over
{2 M_I^4} } \ \ , i\not= j \not= k\not= i  \nonumber \\
\alpha _{5_i} \  &  = & \    {  { \lambda_I M_i^2 } \over
{2 M_I^2 } } \ \ ;\ \ 
\alpha _3 \ =\ { {\lambda _I}\over 2 } \ , 
\label{gaugci}
\eeqa
where $M_i=1/R_i$. From the above formulae we observe that,
unlike what happens in the heterotic case, $M_{Planck}$ and 
$M_I$ do not need to be of the same order of magnitude \cite{witten}. 

Consider  for example the simple isotropic case in which all 
compactification radii are taken equal, $R_i=R=1/M_c$. Then one gets
\beq
M_{Planck}^2 \ =\ {{8M_I^8}\over {\lambda _I^2 M_c^6} } \ ;\ 
\alpha _p \ = \ { {\lambda _I}\over 2} ( { {M_c }\over {M_I} })^{p-3 } \ ,\ 
p=9,7,5,3
\label{isotr}
\eeq
that combined give the following relationship
\beq
{ {M_c^{(p-6)}} \over {M_I^{(p-7)} } } \ =\ { {\alpha _p M_{Planck} }\over 
{\sqrt {2} }  }\ = \ 3.5\times 10^{17} \ GeV \ ,
\label{const}
\eeq
where we have assumed $\alpha _p=\alpha _X\approx 1/24$, i.e., the 
unification value obtained from the extrapolation of low-energy coupling 
constants. Notice that in principle these equations give us a certain 
freedom to play with the values of the Type I string scale $M_I$ and the
compactification scale $M_c$. This is to be compared to the analogous 
equation in the perturbative heterotic case where the relation 
$M_{string}=\sqrt{\frac{\alpha_X}{8}} M_{Planck}$ fixes the value of the 
string scale independently of the compactification scale. Notice also that 
if we want to remain in the Type I weak-coupling regime, due to relations 
(\ref{isotr}, \ref{const}), one should obey the constraint
\beq
\lambda _I\ =\ 2\alpha _p \ ({ { M_I}\over {M_c} })^{p-3} \ = \
2\sqrt{2} { { M_I^4}\over {M_c^3M_{Planck} } }\  \leq  O(1) \ .
\label{pert}
\eeq
Thus we remain in perturbation theory (in this simple isotropic case) only 
if $M_I$ is not much higher than $M_c$.
Notice that if we insist in setting $M_I=M_c$ we get into trouble
because eq.(\ref{const}) would imply $\alpha _p= \sqrt{2}M_I/M_{Planck}$, 
too small a value for identifying $\alpha_p$ with the SM unified gauge 
coupling.

There are a number of natural options for the string scale $M_I$:

{\bf i) $M_I\approx M_{Planck}$}. As we said, this is the option which
is forced upon us in perturbative heterotic vacua \cite{rev}. In this case 
$M_c\approx M_I$ and gauge coupling unification should take place also 
about the same scale.

{\bf ii) $M_I\approx M_X$} \cite{witten}. Here $M_X$ is the GUT scale 
or the scale at which the extrapolated gauge couplings of the minimal 
supersymmetric standard model (MSSM) join. Numerically this is of order 
$10^{16}$ GeV. This corresponds (for the 3-brane case) to choices for 
$M_c$ only slightly below $M_I$.

{\bf iii) $M_I\approx \sqrt{M_WM_{Planck}}$}. This is the geometrical
intermediate scale $\approx 10^{11}$ GeV which coincides with the 
SUSY-breaking scale in models with a hidden sector and gravity mediated 
SUSY breaking in the observable sector. The interest of this choice has 
been recently pointed out in ref.\cite{BIQ} (see also ref.\cite{benakli}). 
In this case one has (in an isotropical 3-branes situation) 
$M_c/M_I\approx 0.01$ and there should be precocious unification of gauge 
coupling constants at $M_X=M_I$.

{\bf iv) $M_I\approx 1$} TeV. This is the 1 TeV string scenario considered in 
refs.\cite{lykken,untev,anton,bajogut,sundrum,shiutye,bursto,kakutev,bursta}. 
In this case it should be (in an isotropical 3-brane situation) 
$M_c/M_I \approx 10^{-5}$. Power-like running
\cite{tv}  of gauge couplings 
 is required \cite{bajogut,gr}
 to get unification at $\approx 1$ TeV.

Many of the results we are going to discuss are independent of the scale
we assume for $M_I$, but we will often have in mind the first three 
possibilities. In particular, in the next subsection we are going to 
consider the possibility of identifying the scale of unification of coupling 
constants $M_X\approx 2\times 10^{16}$ GeV with the string scale 
(possibility ii) above). On the other hand, in section 5, we will consider 
the generation of the $M_W/M_p$ hierarchy and argue that possibility iii) 
with $M_I\approx \sqrt{M_WM_{Planck}}$ allows for a natural understanding of this
hierarchy \cite{BIQ} . This does not require  the existence of 
extra hierarchy-generating mechanisms like gaugino condensation
which seem to be necessary in possibilities i) and ii). The 
effective Lagrangian derived in section 6 as well as the study of soft 
SUSY-breaking terms induced in dilaton/moduli dominated schemes apply
to the first three schemes. The novel mechanism for gauge coupling 
unification based on D-branes sitting close to singularities discussed
in chapter 8 is in principle possible for all four schemes.

\subsection{Gauge coupling unification with an isotropic compact space}
The possibility of better accommodating the scale of gauge coupling 
unification $M_X = 2\times 10^{16}$ GeV within Type I string theory was 
first considered in ref.\cite{witten}. Here we would like to consider this 
issue in a more systematic way for the different possibilities of 
embedding the SM into D-branes. We first discuss the case of an isotropic 
compactification with all $R_i=R=1/M_c$ and with the SM gauge group embedded 
in a single set of coincident p-branes of the same type (same p and same 
world-volume). 

First we should define what are we going to call 
gauge coupling unification scale. We are going to assume 
direct unification of the $SU(3)\times SU(2)\times U(1)$
couplings without an intermediate GUT symmetry. In this context it is 
somewhat confusing what is the meaning of unification scale. For us it is
going to be the scale above which the SM couplings join and 
stop running differently. This may happen because at that 
scale there is the string scale beyond which running no longer makes sense 
and tree-level string constraints between the different couplings apply. 
In this case one has $M_I=M_X$. Alternatively, it may be that the 
couplings unify at a compactification (winding) scale $M_X=M_c (M_w) \leq 
M_I$. Both types of unification may occur depending on how we embed the 
SM into p-branes.

{\it i) Embedding the SM inside 9-branes}

\noindent
In this case $p=9$ and, using (\ref{const}), we have
\beq
{{M_c^3} \over {M_I^2}} \ =\ {{\alpha_9 M_{Planck} }\over {\sqrt{2}}}\ ; \ 
\lambda_I\ =\ 2\alpha_9({{M_I}\over {M_c}})^6 \ .   
\label{nine}
\eeq
Now the open strings have only Neumann boundary conditions in all dimensions
and hence there are KK modes with masses of order $M_c$ but no open string  
windings. In principle one can try to identify the unification scale 
$M_X$ with either $M_c$ or $M_I$. In the first case it must be
$M_c<M_I$ which implies from eq.(\ref{nine}) that $M_X=M_c > 3.5\times 
10^{17}$ GeV, which is too large as unification scale.
The alternative possibility $M_X=M_I < M_c$ is consistent. 
Thus, in order to get $M_X=M_I=2\times 10^{16}$ GeV it is enough to
choose $M_c=5.2\times 10^{16}$, only slightly above the string scale.

\begin{figure}[t]
\begin{center}
\epsfig{file=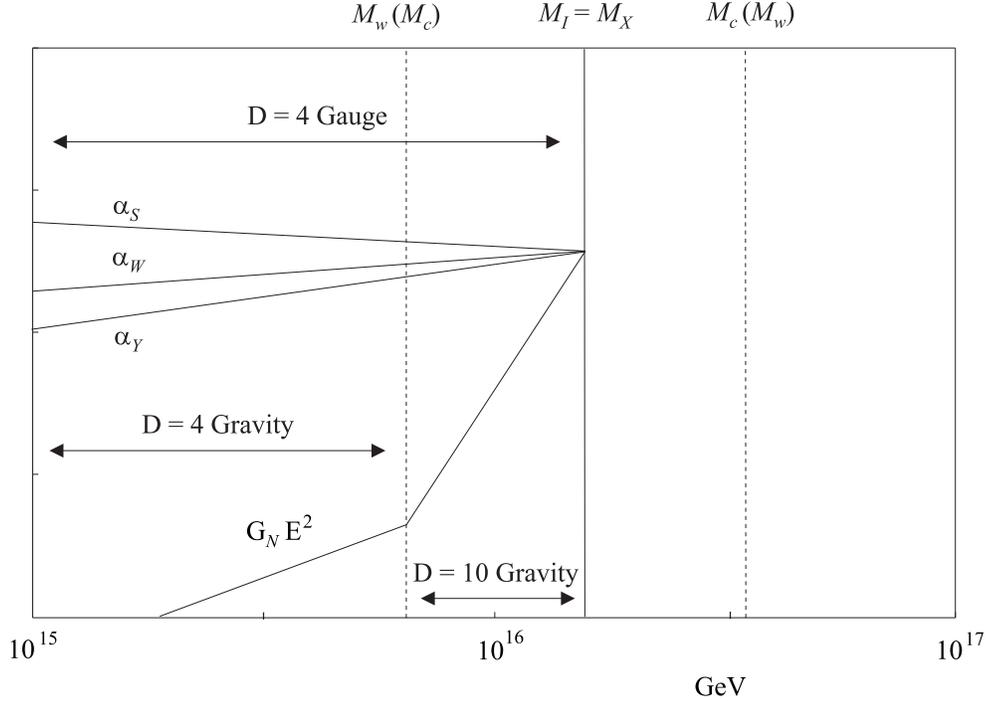,height=9.5cm} 
\end{center}
\caption{\it \small Running of the dimensionless gravitational coupling 
and gauge couplings with energy. The SM is embedded in a $9$($3$)-brane
sector.}
\label{fig1a}
\end{figure}

On the closed string sector there will be both KK and winding modes.
The winding modes will have masses of order $M_w=M_I^2/M_c= 7.7 
\times 10 ^{15}$ GeV. However they will cause no effect in the running 
since they are neutral objects with only gravitational interactions with 
charged matter. In an effective way, the charged SM particles 
embedded in the 9-branes will see only four dimensions
below $M_c=5.2\times 10^{16}$ GeV whereas the closed string
gravitational/moduli sector will see ten dimensions already above 
$M_w= 7.7 \times 10^{15}$ GeV. Thus the unification of gauge and 
gravitational interactions proceeds as schematically shown in 
Fig.\ref{fig1a}. 
\begin{figure}[t]
\begin{center}
\epsfig{file=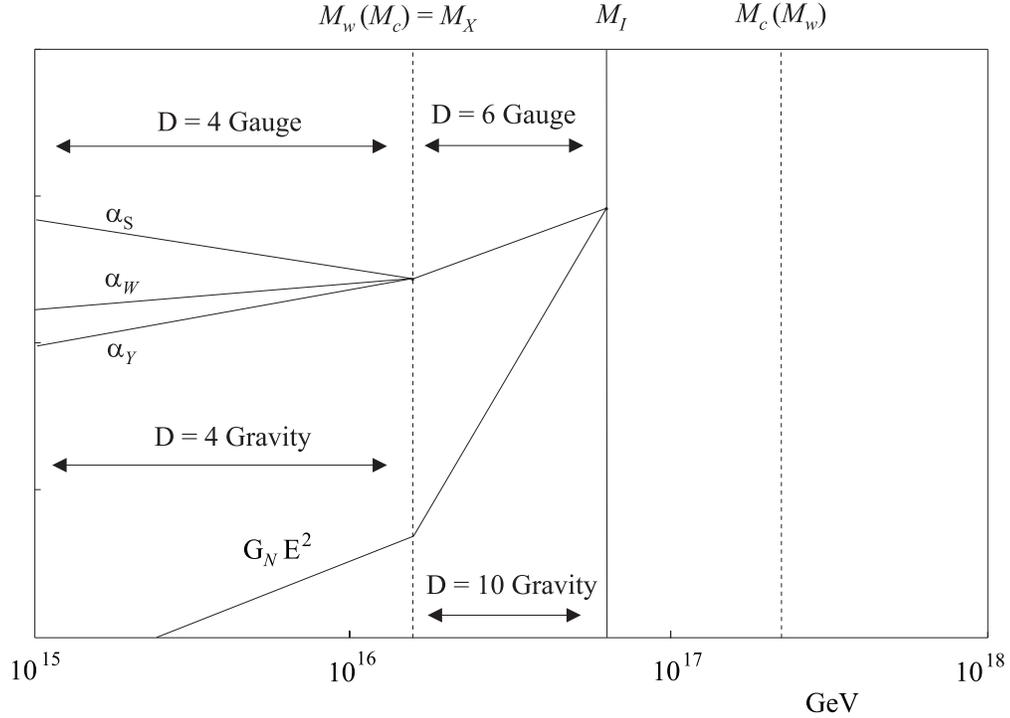,height=9.5cm} 
\end{center}
\caption{\it \small  Running of the dimensionless gravitational coupling 
and gauge couplings with energy. The SM is embedded in a $7$($5$)-brane
sector.}
\label{fig1b}
\end{figure}
Notice that in this scenario the $D=10$ Type I dilaton
remains within the perturbative regime with
$\lambda _I=6.5\times 10^{-3}\alpha _9=2.7 \times 10^{-4}$.
 
{\it ii) Embedding the SM inside 3-branes}

\noindent
In this case $p=3$ and we have
\beq
{{M_I^4} \over {M_c^3 }} \ =\ { {\alpha_3 M_{Planck} }\over {\sqrt{2}}}\ ;\ 
\lambda_I\ =\ 2 \alpha_3 \ .
\label{three}
\eeq
This configuration is related to the one above by a T-duality 
transformation in all six compact dimensions \cite{anton,lpt,shiutye}.
Now open strings have Dirichlet boundary conditions in all
six compact dimensions and hence have winding modes with
masses of order $M_w=M_I^2/M_c$ but no KK modes. On the other
hand eq.(\ref{three})  tells us that $M_I>M_c$, otherwise 
$M_I>3.5\times 10^{17}$ GeV which is too large for unification.
Thus necessarily $M_c<M_I$. One could naively say that in this case
one should identify the unification scale $M_X$ with $M_c$.
However this can not be because there are no open string charged KK 
states with masses of order $M_c$ but only winding 
states with masses $M_w=M_I^2/M_c$, higher than $M_I$. Thus one has
actually to identify $M_X=M_I=2\times 10^{16}$ GeV, just like in
the 9-brane case. In fact the physics of both cases is exactly the 
same, just exchanging KK modes by windings. One gets consistency 
with eq.(\ref{three}) by choosing $M_c=7.7 \times 10^{15}$ GeV, $M_I=2 
\times 10^{16}$ GeV. This is schematically shown in Fig.\ref{fig1a}. 
As it happens with 9-branes, although above $M_c$ gravity lives in ten 
dimensions, the SM lives on the world-volume of the 3-branes and only 
feels four dimensions up to the winding scale $M_w$. The Type I dilaton 
remains in the perturbative regime with $\lambda_I = 2\alpha_3$.

This is qualitatively similar to the Witten scheme
\cite{witten} for gauge coupling unification inside strongly coupled 
$E_8\times E_8$ heterotic. However, in the present case, there is
an intermediate field theory regime with $D=10$ (instead of $D=5$) 
dimensions.

{\it iii) Embedding the SM within 7-branes}

\noindent
Let us consider a set of $7_i$-branes with their world-volume 
spanning Minkowski space plus the two complex compact dimensions
orthogonal to the $X_i$-th complex plane. We still assume for the 
moment an isotropic compactification of scale $M_c$.
In this case (\ref{isotr}) and (\ref{const}) give
\beq
M_c  \ =\ { {\alpha _7 M_{Planck} }\over {\sqrt{2} } }\ ;\ \lambda _I\ =\
         2 \alpha _7({ {M_I}\over {M_c}})^4  \ ,
\label{siete}
\eeq
so that, unlike previous cases, $M_c=3.5\times 10^{17}$ GeV is fixed and
appears to be too large to be identified with the unification scale
$M_X$. However in the present case open strings have Dirichlet boundary 
conditions on the $X_i$-th complex compact dimension and Neumann boundary 
conditions on the other two complex planes. Thus there will be charged 
winding modes with masses $M_w=M_I^2/M_c$ along the $i-th$ direction and
KK modes with masses $M_c$ along the other two. Thus we have the hierarchy 
of scales $M_w<M_I<M_c$. Since now the winding modes are in general charged 
with respect to the SM interactions we must identify the unification scale 
$M_X$ with the mass of the winding modes, $M_X=M_w=M_I^2/M_c=2\times 
10^{16}$ GeV. Thus one obtains $M_X=2\times 10^{16}$ GeV by taking 
$M_I=8.3\times 10^{16}$ GeV. The structure of scales is schematically 
shown in Fig.\ref{fig1b}.

In the present case the unification scale would coincide with the scale 
$M_w$ at which a new threshold appears corresponding to have 
effectively six dimensions felt by the charged fields (ten dimensions 
are felt by the bulk gravity fields). This interpretation is clearer 
in the T-dual 5-brane scheme which we discuss below. One has for the 
Type I dilaton coupling
\beq
\lambda_I \ = \ 2 \alpha_7 ({{M_I}\over {M_c}})^4 \ = \ 
              6.3 \times 10^{-3} \ \alpha_7 
\label{nopersiete}
\eeq
thus for a realistic coupling, $\alpha _7=1/24$, we are well within the 
perturbative Type I regime.

{\it iv) Embedding the SM within 5-branes}

\noindent
In this case we have
\beq
{{M_I^2} \over {M_c }} \ =\ {{\alpha _5 M_{Planck}}\over {\sqrt{2}}}\ ;
      \ \lambda _I\ =\ 2 \alpha _5({{M_I}\over {M_c}})^2  \ .
\label{cinco}
\eeq
Let us consider one set of $5_i$-branes whose world-volume spans Minkowski 
space plus the $i$-th compact plane. Open strings will have Dirichlet 
boundary conditions on the two complex planes transverse to the $i$-th plane 
and Neumann boundary conditions in the other dimension.
Thus there will be charged KK modes with masses $M_c$ along the
$i$-th compact plane and winding modes with masses
$M_w=M_I^2/M_c$ along the other two. From eq. (\ref{cinco}) the winding mass 
is fixed to be $M_w = 3.5 \times 10^{17}$. So in this case we must 
identify the unification scale with  the compactification
scale $M_X=M_c=2\times 10^{16}$ GeV. The physical scales are
much the same as the ones found for the 7-brane case. Thus 
correct unification is obtained by setting $M_I=8.3\times 10^{16}$ GeV.
This is schematically shown in Fig.\ref{fig1b}.
However, unlike the previous 7-brane case one finds for the
Type I dilaton coupling
\beq
\lambda_I \ =\ 2 \alpha_5 ({{M_I}\over {M_c} })^2\ =\ 35 \alpha_5 \ =\ 1.5 \ .
\label{nopercinco}
\eeq
Thus in this case  we are at the border of the  range of validity of the 
theory. One must not be surprised that in the present 5-brane case one 
has a larger Type I coupling  whereas in the T-dual with 7-branes
is much smaller (\ref{nopersiete}). This is because under T-duality  in all compact
dimensions $\lambda_I$ transforms like $\lambda_I \rightarrow \lambda_I 
(M_c/M_I )^6$ (see eq.(\ref{rescaleo})).

As a general conclusion one observes that assuming unification
of SM interactions within a single type of p-brane, and 
assuming the simplest strictly {\it isotropic} compactification,
one can get consistent results for all p-branes, with
$p=9,3,7,5$. In the case of 9- and 3-branes one identifies
the unification scale with the string scale. In the case of
7- and 5-branes one must identify it with a winding or 
KK threshold. In all cases one remains within Type I perturbation 
theory except  in the case of 5-branes in which one sits close to the 
non-perturbative regime.

\subsection{Non-isotropic compactifications}

We will not attempt to  make a general analysis of all possibilities.
We just want to show here how departure from strict isotropy opens the
way to new possibilities. Thus e.g. one can consider 9-branes (or 
3-branes) scenarios in which the unification scale has to be identified with 
a compactification KK (winding) scale and not with the string scale $M_I$.
Also, the 7-brane or 5-brane scenarios with isotropic compactifications
considered in previous section have T-dual equivalents in which they are
realized as non-isotropic compactifications within 9-branes or 3-branes.
Other anisotropic scenarios in which different types of branes
appear are described in the next subsection.

{\it i) Anisotropic 9(3)-brane scenarios with a KK(winding) unification scale}

\noindent
We saw in previous section how unification of couplings 
within a 9(3)-brane with an isotropic compact space leads naturally
to identify the unification scale $M_X$ with the string scale $M_I$.
We will consider now a non-isotropic situation with mass scale
$M_2$ in the first complex plane and mass scale $M_4$ for the
other four compact dimensions. For 9-branes, using (\ref{planck}) and
(\ref{gaugci}), we will have now
\beq
{{M_2 M_4^2} \over {M_I^2 }} \ =\ {{\alpha_9 M_{Planck} }\over{\sqrt{2}}}\ ;\ 
\lambda_I\ =\ 2 \alpha_9({{M_I^6}\over {M_2^2 M_4^4}}) \ .
\label{nineunis}
\eeq
A consistent solution to these equations is obtained identifying $M_X=M_I=2 
\times 10^{16}$ GeV and $M_2=M_4=M_c= 5.2\times 10^{16}$ GeV. This is the 
isotropic situation discussed above. Alternatively one can consider a 
non-isotropic hierarchy of scales $M_2<M_I<M_4$ and identify the unification 
scale with the $M_2$ KK scale, $M_X=M_2=2\times 10^{16}$ GeV. In this case 
one has to put $M_4^2=17.5 M_I^2$ and the value of $M_I$ is only constrained 
above by the perturbative condition $\lambda_I<1$. 
One can obtain a physically similar scale structure by embedding the SM within
3-branes by making a T-duality transformation over all the six compact 
dimensions. The structure is the same but one has to replace the scales $M_2$ 
and $M_4$ respectively by winding scales $M_I^2/M_2$ and $M_I^2/M_4$.
 
{\it ii) Anisotropic 9(3)-brane embeddings dual to 5(7)-brane isotropic 
embeddings}

\noindent
We saw how we can embed the SM couplings within a set of 5-branes with
an isotropic compactification of size $M_c$. In that case we had $M_X=M_c
=2\times 10^{16}$ GeV $ <M_I<M_w$. A similar structure of physical
scales can be obtained embedding the SM inside, say, 9-branes with
an {\it anisotropic} compactification with $M_2=M_X$ and $M_4=M_w$.
In fact both scenarios are related by a T-duality transformation
along the four compact dimensions with mass scale $M_4$.
However in the 9-brane case one finds that the Type I coupling is as small 
as $\lambda_I=2 \alpha_9 M_I^6/(M_2^2 M_4^4)=0.11\alpha_9$, well
within the perturbative regime.

Something analogous can be said of the isotropic schemes with
the SM embedded into 7-branes. An anisotropic compactification
with the SM embedded into 3-branes can be found which gives
rise to the same pattern of scales and perturbative unification.

\subsection{Embedding $SU(3)$ and $SU(2)$ within different sets of
p-branes}

It is a plausible situation to assume that the $SU(3)$ and the 
$SU(2)$ groups of the SM could come from different sets of
p-branes \cite{ibanez,shiutye}. By different sets we mean p-branes 
whose world-volume is not identical. In particular, notice that the SM 
contains particles (the left-handed quarks) transforming 
both under $SU(3)$ and $SU(2)$. That means that there must be
some overlap of the world-volumes of both sets of p-branes.
Thus e.g., one cannot put $SU(3)$ inside a set of 3-branes 
and $SU(2)$ within another set of parallel 3-branes on a different point
of the compact space since then there would be no massless
modes corresponding to the exchange of open strings between 
both sets of brane which could give rise to the left-handed
quarks.  Thus we need to embed $SU(3)$ inside a p-brane
and $SU(2)$ within a q-brane in such a way that their
corresponding world-volumes have some overlap.

Let us first consider what is the relationship between coupling constants 
for gauge groups coming from p-,q-branes. From eq.(\ref{isotr}) one has 
for an isotropic compactification with compact scale $M_c$:
\beq
\alpha_p\ =\ \alpha_q ({{M_I}\over {M_c} })^{q-p}
\label{relalfas}
\eeq
Now, if we want to preserve $N=1$ SUSY one has $q-p=0,\pm 4 $.
Thus, for an isotropic compactification one naturally has $\alpha_{7_i} 
= \alpha_{7_j}$ and $\alpha_{5_i}=\alpha_{5_j}$ whereas $\alpha_9 = 
\alpha_{5_i} (M_c/M_I)^4$ and $\alpha_3=\alpha_{7_i} (M_I/M_c)^4$.
In order to embed the $SU(3)$ and $SU(2)$ gauge factors inside different 
brane sectors the possibilities are (p,q) = $(9,5_i)$, $(7_i,7_j)$, 
$(5_i,5_j)$, $(3,7_i)$ with $i\not=j$. Thus one naturally obtains 
$\alpha_S=\alpha_W$ for the cases (p,q) = $(7_i,7_j)$, $(5_i,5_j)$.
In the other two cases one will have $\alpha_S/\alpha_W=(M_c/M_I)^{\pm 4}$ 
and thus one of the couplings will be suppressed compared to the other one 
unless $M_c=M_I$. But in such a case, due to (\ref{const}), one is forced 
to have $M_I=3.5\times 10^{17}$ GeV, too large a value for unification.

In fact if one goes beyond the isotropic scale, a certain 
amount of fine-tuning is required in all cases in order to obtain 
$\alpha_S=\alpha_W$. Indeed, if the compact scale in the i-th direction 
is denoted by $M_i$ one has in this more general case (see eq.(\ref{gaugci})):
\beqa
\alpha_{5_i}\ = \ \alpha_{5_j} ({ {M_i}\over {M_j} })^2 \  & ; & 
\alpha_{7_i}\ = \ \alpha_{7_j} ({ {M_j}\over {M_i} })^2  \nn  \\
\alpha_{7_i}\ = \ \alpha_{3} ({ {M_jM_k}\over {M_I^2} })^2  & ; &
\alpha_{9}\ = \ \alpha_{5_i} ({ {M_jM_k}\over {M_I^2} })^2 
\label{alfillas}
\eeqa
where $i\not=j\not=k\not=i$. Thus even in the cases (p,q) = $(7_i,7_j)$, 
$(5_i,5_j)$ one has to adjust $M_i=M_j$ to a good precision in order to have 
$\alpha_S=\alpha_W$.

On the other hand one would in principle expect that the embedding into 
different sets of p-branes would allow for new gauge coupling unification 
possibilities \cite{shiutye} if one is prepared to accept some degree of 
fine-tuning of the compactification scales. 
For example  one can consider the $SU(3)$ and $SU(2)$ gauge couplings 
crossing  at $M_X=2\times 10^{16}$ GeV and then diverging at higher 
energies. One could have $M_i\approx M_I=3.5\times 10^{17}$ GeV.
If $SU(3)$ and $SU(2)$ come from  different sets of 7-branes $7_i$, $7_j$, 
a small difference between $M_i$ and $M_j$ is able to explain the fact that 
at $M_I$ the running gives $\alpha_W > \alpha_S$. Something analogous may 
be done in the $(5_i,5_j)$ scheme. However there are some difficulties 
with the unification of the hypercharge $U(1)_Y$, as we now explain.

In Type IIB orientifold models, and in general on the world-volume of 
D-branes, $SU(n)$ groups come along with a $U(1)$ factor so that indeed 
we are dealing with $U(n)$ groups in which both $SU(n)$ and $U(1)$ share 
the same coupling constant. Assume that the $U(1)_Y$ is embedded into 
$U(3)\times U(2)$ in such a way that the left-handed quarks (belonging to 
the p-(q-)brane sector) have the standard hypercharge assignment. Then one
obtains:
\beq
{1\over {\alpha_Y} }\ =\ {2\over 3} {1\over {\alpha_S}} \ + \ 
{1\over {\alpha_W } } 
\label{wangle}
\eeq
In particular one has the expression:
\beq
\sin^2\theta _W \ =\ { {3/8}\over {3/4+\alpha_W/4\alpha_S}}
\label{weinb}
\eeq
Thus for $\alpha_S=\alpha_W$ one obtains the standard GUT normalization 
for couplings, $\alpha_Y = 3/5 \alpha_W$ and $\sin^2 \theta_W = 3/8$.

Notice that for $\alpha_W > \alpha_S$ (as happens if the couplings keep on 
running beyond $M_X=2\times 10^{16}$ GeV), one has $\sin^2\theta_W < 3/8$.
This goes in the wrong direction for unification of the three coupling 
constants (notice that at $M_I>M_X$ it should be $5/3\alpha_Y > \alpha_W$).
Thus, even if we adjust appropriately the $M_i$ in order to match $\alpha_W
>\alpha_S$ at $M_I=3.5\times 10^{17}$ GeV, the value of $\sin^2\theta_W$
at that scale is going to be too small to be consistent with low-energy
data. One can consider the possibility of $U(1)_Y$ not being completely
included in $U(3)\times U(2)$ but having some component in other 
brane sector. In such a case the equality sign in eq.(\ref{wangle}) 
should be replaced by an inequality $>$, leading to an even smaller 
weak angle. 

We thus conclude that embedding $SU(3)$ and $SU(2)$ inside
different type of p-branes and trying to maintain $M_I\approx M_i\approx
3.5\times 10^{17}$ GeV do not lead to consistent low-energy results for
$\sin^2\theta_W$, even accepting some degree of fine-tuning.
On the other hand the appropriate GUT relationships $\alpha_S=\alpha_W = 
5/3\alpha_Y$ are obtained if we embed $SU(3)$ and $SU(2)$ into either 
a set of $(5_i,5_j)$ or $(7_i,7_j)$ branes with all $M_i=M_c=2\times 
10^{16}$ GeV in the first case and $M_I^2/M_c=2\times 10^{16}$ GeV in 
the second. In that case the structure of mass scales would be analogous 
to those we found for the case of embedding the SM within a 
single $5$-brane or $7$-brane sector, choosing $M_I=8\times 10^{16}$
GeV. The two cases are however not equivalent since the world-volumes
of the $5$-branes or $7$-branes in which $SU(3)$ and $SU(2)$
live in the present case are now different.
%
%
\section{Micro gauge/Yukawa interactions}
%
%
Eq.(\ref{alfillas}) shows that there may potentially appear gauge
couplings widely different if there are large differences  between
the compactification scales and/or string scale. Consider e.g.,
a general situation with one set of 3-branes plus sets of
$7_i$-branes of the different types $i=1,2,3$. Then the corresponding
gauge couplings will be related by:
\beq
g _{7_i}\ = \  g _{3} ({ {M_jM_k}\over {M_I^2} })  \  ; \
g _{7_i}\ = \   g_{7_j}  ({ {M_j}\over {M_i} }) \ .   
\label{acoplillos}
\eeq
If all scales $M_i$ are of the same order of magnitude and all not
far away from the string scale $M_I$, as mostly considered in the previous
chapter, the different gauge couplings will have similar sizes.
However in schemes with large compactification radii very small gauge 
couplings  naturally appear \cite{shiutye}.

Consider first an isotropic case with $M_i=M_c$. Depending on our choice
for the string scale one has:

{\bf  i) $ M_I=1$ } TeV
 
 If one assumes
$M_c\approx 10 MeV $ as in isotropic schemes with $M_I=1$ TeV, one gets
$g_{7_i}=10^{-10}g_3$. Thus in a scheme with a 1 TeV string scale and the
SM sitting on 3-branes, the gauge interactions coming from  7-branes
would  appear to be  very much suppressed like $(M_c/M_I)^4$. They may be 
even more suppressed in a non-isotropic scenario with two dimensions with 
inverse size of order $10^{-3}$ eV and the other two slightly below 1 TeV
\cite{untev,anton}. In this case eq.(\ref{acoplillos}) yields 
$g_{7_i}=10^{-15}g_3$. There is however a generic problem concerning the 
possibility of these micro-gauge interactions  in $M_I=1$ TeV scenarios. 
Indeed, if the compactification scales are as low as 10 MeV or even smaller, 
the 7-branes wrapping around those compact dimensions will give rise to 
open string KK modes which (unlike the closed string ones) will in general 
couple to the visible 3-brane sector. Thus, unless  one finds a mechanism 
to hide these 7-brane sectors and to forbid the presence of
massless chiral fields in the (3-7) open string sector, this   
micro gauge interactions do not seem viable.

More generally, having 7-branes along with 3-branes present in 
a 1 TeV scenario will in general be dangerous unless the 
four compact dimensions around which the 7-branes are wrapping 
are close to the TeV scale. In this way the open string KK states 
will be sufficiently heavy. These requires the other two 
compact dimensions to have a scale of order $10^{-3}$ eV
in order to reproduce the usual Planck mass. But then the
gauge interactions associated to these 7-branes are not going
to be suppressed since they have couplings proportional to the
compact scales of wrapped dimensions. 
In conclusion, in 1 TeV schemes, micro-gauge interactions due to 
7-branes wrapping on large dimensions do not seem viable. This is 
unfortunate because in these 1 TeV schemes the stability of the proton 
is particularly problematic. It has been proposed \cite{untev,anton}
that some $U(1)_B$ field living in the bulk and gauging baryon
number could forbid fast proton decay. On the other hand it is not 
obvious how to get such bulk $U(1)$'s in the orientifold scheme. A 
natural alternative would  have been a micro-gauged $U(1)_B$ living on 
a 7-brane with its world-volume including two large compact dimensions. 
We have just seen this does not look viable.

{\bf ii)  $M_I=10^{11}$ } GeV

This is the case studied in ref.\cite{BIQ} (see also next section).
In this case one has $M_c=10^9$ GeV and one gets
$g_{7_i}=10^{-4}g_3$. Thus if the SM is contained in a 3-brane
sector, the gauge interactions on the 7-brane world-volume
will be of the typical size of first generation Yukawa couplings.
Smaller gauge couplings may be obtained in non-isotropic 
configurations. Indeed if e.g. the compact dimensions 
$X_{2,3}$ have sizes $M_2\approx M_3\approx M_I$ one can have 
$M_1$ as small as $10^5$ GeV and still maintain the correct Planck mass.
Then we would get micro-gauge interactions for gauge groups 
in the $7_2$ and $7_3$ world-volumes with $g_{7_{2,3}}=10^{-6} g_3$.  

{\bf iii) $M_I=2\times 10^{16}$ } GeV

This is the case put forward in ref.\cite{witten}. In this case
one has $M_c\approx 8\times 10^{15}$ GeV and $g_{7_i}\approx g_3/4$. 
Even in this case the gauge couplings of gauge interactions
coming from the 7-brane sectors are smaller than the 3-brane ones.
 
Notice that the different type of branes map under Type I/heterotic 
duality  into perturbative and non-perturbative gauge groups. Thus, 
from the heterotic point of view, what we find is that the e.g., 
non-perturbative gauge couplings should be much smaller than the 
perturbative ones as long as there are some dimensions which are 
relatively large, like in all three schemes reviewed above.

These micro-gauge interactions could have some interesting applications 
in order to generate some flavor structure in specific models. 
In addition to the micro gauge interactions there will be in general 
accompanying micro-Yukawa couplings of the same size. It is tempting
the idea of associating the first (and perhaps the second) 
SM  generation Yukawa couplings to these micro-Yukawa couplings.
This idea is however difficult to realize due to the micro-gauge
interactions which come along which would give rise to
departures from universality in the SM gauge interactions. More 
concrete semi-realistic models would be needed to test this idea.
%
%
\section{The generation of the $M_W/M_{Planck}$ hierarchy }
%
%
Given the new possibilities for mass scales present in
$D=4$, $N=1$ Type I  string vacua, it is worth revisiting
also the SUSY-breaking scales as well as the generation
of the $M_W-M_{Planck}$ hierarchy.

The orthodox  scenario for the generation of the
$M_W/M_{Planck}$ hierarchy in perturbative heterotic vacua
is  the assumption of the existence of a
hidden sector in the theory which couples only
gravitationally to the usual particles of the SM.
Indeed in perturbative heterotic vacua there are
often this kind of hidden sectors which usually also involve
gauge interactions \cite{rev}. A natural possibility is to assume that
the gauginos of the hidden sector may condense at a
scale $\Lambda \approx 10^{13}$ GeV so that SUSY is
broken and the gravitino gets a mass $m _{3/2}\approx
\Lambda ^3/M_{Planck}^2\approx M_W$.

The existence of hidden sectors is equally natural
within Type I vacua. For example, one can consider
two separate sets of parallel p-branes ($p < 9$) , one set
containing the SM  and the other the hidden sector interactions.
If the positions of these p-branes in the transverse space
are different, there will be no massless states charged
under both p-brane groups. 
In the case of 9-branes the same effect is obtained by adding
discrete Wilson lines to the model. In what follows we will consider
an ideal situation of the above type in which we have
a set of p-branes with unbroken $N=1$ SUSY containing the
SM  and a distant (hidden) set of parallel p-branes in which
SUSY is somehow broken and one has $N=0$. Now, the SM
does not feel directly the breaking of SUSY taking place in the
hidden p-brane sector. The effects of SUSY-breaking can only be
transmitted  through the closed string sector fields, which are
the only ones that can move into the bulk and couple to
both p-brane sectors. Thus the SUSY breaking felt by the
SM fields will be suppressed by $M_{Planck}$ powers. In
particular one expects that the soft SUSY-breaking terms
felt in this visible sector will be of order:
\beq
 M_W \ \approx \ m_{3/2} \ \approx \ {{F}\over {M_{Planck} } }
\ \approx \ { {\xi M_{ss}^2}\over {M_{Planck}} }\ ,
\label{debil}
\eeq
where $F$ denotes the SUSY-breaking auxiliary field and
$M_{ss}$ is the relevant physical  scale in the
hidden sector.  Here $\xi $ is just a dimensionless fudge factor whose
size we will discuss below.  It may be of order one or
hierarchically small depending on how the $N=0$
hidden theory is originated.
Combining (\ref{debil}) with (\ref{const}) one has:
\beq
{{M_W}\over {M_{Planck} }  }\ =\
\xi {{\alpha_p^2}\over {2} }\ M_{ss}^2 \frac{M_I^{2(p-7)}}{M_c^{2(p-6)}} \ ,
\label{jerardo}
\eeq
with $p$=9,7,5,3.
Depending on the relative values of $M_I$ and $M_c$ and
which type of p-brane is being considered, the relevant
physical scale in the hidden sector should be identified
with a)$ M_{ss}=M_I$ , b) $M_{ss}=M_c$ or c) $M_{ss}=M_I^2/M_c$
(a winding scale)\footnote{We will not impose for
the moment in this discussion  the constraints
from gauge coupling unification.}.

{\it i) Case with $M_{ss}=M_I$ }

\noindent
Let us start by assuming the case in which the relevant
scale of the hidden sector is $M_{ss}=M_I$. Now, let us consider for 
simplicity an overall compactification scale $M_c$. 
%
%
Using (\ref{jerardo}) one obtains:
\beq
{{M_W}\over {M_{Planck} }  }\ =\
\xi {{\alpha_p^2}\over {2} }\ ({{M_I}\over {M_c} })^{2(p-6)} \ .
\label{jerar2}
\eeq
This equation is remarkable since it shows us that it is
in principle possible to generate a $M_W-M_{Planck}$
hierarchy even for $\xi \approx 1$ without invoking
any hierarchically suppressed
non-perturbative effect like e.g., gaugino condensation.
This is the possibility discussed in \cite{BIQ}.
Indeed, consider first the $p=3$ case. In this case one has
\beq
{ {M_W}\over {M_{Planck} }  }\ =\
\xi { {\alpha _3^2}\over {2} }\ ({ {M_c}\over {M_I} })^{6}
\label{jerar3}
\eeq
If one has $M_I/M_c\approx 160 $ and $\alpha _3=1/24$ one
indeed obtains the desired hierarchy $M_W/M_{Planck}
=10^{-16}$ even for $\xi \approx 1$. This small ratio
arises from a combination of geometrical factors which
amplify the modest input hierarchy  $M_I/M_c$.
The final structure of scales would be
$M_c:M_I:M_I^2/M_c=10^{9}:10^{11}:10^{13}$ GeV.
 The
argumentation works equally well for the $p=9$ case
in which a similar formula is obtained with the
replacements $\alpha _3\leftrightarrow \alpha _9$ and
$M_I/M_c\leftrightarrow M_c/M_I$. In this case the
input hierarchy would be the inverse,
$M_c/M_I \approx 160$.

For this scheme to work one has to be sure that we remain within
Type I perturbation theory. This requires
$\lambda _I = 2\alpha _p(M_I/M_c)^{(p-3)} <1$. This is the case indeed for
3-branes and 9-branes.

In the  case of 7-branes (and their T-duals the 5-branes) the situation is a bit
different. For both cases remaining in perturbation theory requires
$M_I\leq M_c$. Then the winding scale $M_I^2/M_c$ will actually be
lighter than the string scale $M_I$ so we rather take $M_{ss}=M_I^2/M_c$ 
instead of $M_{ss}=M_I$. Thus we turn now to this possibility.

{\it ii)  Case with $M_{ss}=M_I^2/M_c $  }

\noindent
This case corresponds to a hidden p-brane sector in which the
lightest cut-off scale correspond to charged winding modes 
with masses $M_I^2/M_c$ rather than the string excitations
with masses of order $M_I$. Notice that this may be the case only for 
$p=3,5,7$ but not for 9-branes which have no charged winding modes.
Now one parametrizes $F=\xi (M_I^2/M_c)^2 $ and using (\ref{jerardo})
obtains:
\beq
{ {M_W}\over {M_{Planck} }  }\ =\
\xi { {\alpha _p^2}\over {2} }\ ({ {M_I}\over {M_c} })^{2(p-5)}
\label{jerar4}
\eeq
Note that no hierarchy is obtained for 5-branes
(for $\xi \approx 1$). Furthermore, in the case of 3-branes
obtaining a suppression would require $M_c<<M_I$ in which case
one has $M_I<<M_I^2/M_c$ and we rather identify $M_{ss}=M_I$,
as in the previous paragraph. Finally, one naturally obtains
a suppression of $M_W$ versus $M_{Planck}$ in the case of
7-branes. By choosing $M_I/M_c\approx 10^{-3}$ one obtains the
appropriate hierarchy even for $\xi \approx 1$. In the case
of 7-branes one has $M_c=\alpha _7M_{Planck}/\sqrt{2}$ hence we
finally would have scales of order
$M_I^2/M_c : M_I : M_c=10^{11}:10^{14}:10^{17}$ GeV.

{\it iii) Case with $M_{ss}=M_c$ }

\noindent
This case corresponds to a hidden p-brane sector in which the
lightest cut-off scale correspond to KK modes
with masses $M_c$ rather than the string excitations
with masses of order $M_I$. Notice that this may be  
the case only for $p=5,7,9$ but not for
3-branes which have no charged  KK  modes.
We now have $F=\xi M_c^2$ and using (\ref{jerardo}) one gets:   
\beq
{ {M_W}\over {M_{Planck} }  }\ =\
\xi { {\alpha _p^2}\over {2} }\ ({ {M_I}\over {M_c} })^{2(p-7)}
\label{jerar}
\eeq
Note that no hierarchy is obtained for 7-branes
(for $\xi \approx 1 $). Furthermore  in the case of
9-branes obtaining a suppression would require $M_I<<M_c$.
But in that case we rather identify $M_{ss}=M_I$ as in the
first paragraph. Finally, one  naturally obtains
a suppression of $M_W$ versus $M_{Planck}$ in the case of
5-branes. By choosing $M_I/M_c\approx 10^{3}$ one obtains the
appropriate hierarchy even for $\xi \approx 1$. This scheme is really
T-dual to the previous  one and  we
finally would have scales of order 
$M_c : M_I :M_I^2/ M_c=10^{11}:10^{14}:10^{17}$ GeV.

In summary, we have shown that in Type I, $D=4$ vacua the
hierarchy problem is ameliorated by the presence of
geometrical factors of the form $\alpha ^2(M_I/M_c)^{\pm 6}$
or $\alpha ^2(M_I/M_c)^{\pm 4}$
depending on the cases.
These geometrical factors may in fact explain the whole
of the $M_W/M_{Planck}$ hierarchy ($\xi =1 $) in terms
of modest initial hierarchies $M_I/M_c\approx 10^{\pm 2}$.

If all of the hierarchy is explained in that way,
one has to give up standard  gauge coupling unification
at $2\times 10^{16}$ GeV as 
discussed in section 3 . Indeed, in principle the
mass scale $M_{ss}$ discussed here corresponds to the
scale one expects  coupling constants to join, thus
one would expect $M_X=M_{ss}$. Thus 
e.g., if one takes $M_I=\sqrt{M_WM_{Planck}}$ and  one assumes
$\xi \approx 1 $ the unification scale should be identified
with $M_{ss}\approx 10^{11}$ GeV.
In ref.\cite{BIQ} it was shown how the addition of some extra 
charged matter fields beyond those of the MSSM can give
rise to precocious gauge coupling unification at
$10^{11}$ GeV. Furthermore we will show in
section 8  a novel mechanism by which gauge coupling 
unification can be understood  if the observable 3-branes 
sit close to a repaired singularity. 

\subsection{Gaugino condensation}

A natural possibility often considered to generate a hierarchically small
$\xi $ is gaugino condensation. Let us assume that there
is gaugino condensation in the hidden p-brane sector.
In this case one expects the generation of an auxiliary
field:
\beq
F\ =\ { { \Lambda ^3}\over {M_{Planck} } }\ \approx
e^{-3/2\beta \alpha _p}\times { {M_{ss}^3}\over {M_{Planck} } }
\label{efe}
\eeq
and hence comparing with the formulae in previous section one has
\beq
\xi \ =\ e^{-3/2\beta \alpha _p}\times { {M_{ss}}\over {M_{Planck} } } \ .
\label{cha}
\eeq
Here $\beta$ is the beta-function coefficient  of the
condensing gauge group which has coupling $\alpha _p$.
Coming back to the three possibilities considered above   
for the value of the cut-off scale $M_{ss}$ i.e.,
$M_{ss}=M_I, M_I^2/M_c, M_c$ one obtains
\beq
\xi \ =\  e^{-3/2\beta \alpha _p}{ {\alpha _p}\over {\sqrt{2} } }   
({ { M_I}\over {M_c} })^{(p-a)}
\label{che}
\eeq
with $a=6,7,5$ respectively for the three $M_{ss}$
possibilities. Thus, for example for the case of
3-branes which have $M_{ss}=M_I$, one gets a hierarchy:
\beq
{ {M_W}\over {M_{Planck} }  }\ =\   
e^{-3/2\beta \alpha _3} { {\alpha _3^3}\over {2\sqrt{2}} }\ ({ {M_c}\over
  {M_I} })^{9} \ . 
\label{jerarg3}
\eeq

One can now simultaneously impose correct gauge coupling unification
in the visible 3-brane sector with $M_I=M_{ss}=2\times 10^{16}$ GeV,
as we described in section 3. In this case one has $M_c=7.7\times
10^{15}$
GeV so that, for $\alpha _3=1/24$ one gets
$M_W/M_{Planck}= 10^{-8}\times \exp(-3/2\beta \alpha _3)$.

%
%
\section{Effective low-energy Lagrangian in $D=4$ $N=1$ compact 
orientifold models}
%
%
%
In this section we will extract the structure of the K\"ahler 
potential corresponding to the untwisted closed string fields
(dilaton and moduli) as well as that of the charged chiral fields
coming from the different p-brane sectors in Type IIB 
orientifold models. Likewise we will discuss the gauge kinetic functions
and superpotential of this type of models.

As we described in section 2, we have two types of massless
$N=1$ chiral fields in these models. 

{\it i) Closed string chiral singlets}

\noindent 
These are the chiral singlets coming from the {\it closed string} 
sector of the theory. Their real parts come from the NS-NS sector
and the imaginary part from R-R fields.
These will include the complex dilaton 
field $S$ and the untwisted moduli fields $T_i$, {\it i}=1,2,3,
one per complex compact dimension\footnote{ In $Z_3$ and $Z_6$ there are 
additional off-diagonal moduli. In the case of $Z_6'$ there is 
also one complex structure modulus field $U$. We will
ignore these particularities here.}.
In addition there are singlets fields $M_a$ associated to 
the different fixed point singularities in the underlying 
orbifold. These are more model dependent and we will 
describe some important aspects of their couplings in
chapter 8  but
for the moment we will ignore them.

{\it ii) Charged open string chiral fields}

\noindent 
The gauge groups and charged chiral fields will depend on the
type and number of Dp-branes present in the vacuum as well
as their location in the compact coordinates. One can consider, as the 
most general class of Type IIB toroidal orientifolds, the case with
one set of 9-branes and three sets of 5-branes, $5_i$, {\it i}=1,2,3 
with world-volumes spanning the {\it i}-th compact complex dimension
and Minkowski space. This is equivalent to say that the 
$5_i$-brane is wrapping around the {\it i}-th torus.
Other possible $N=1$ configurations are equivalent to this under T-duality 
transformations along different combinations of compact complex 
dimensions\footnote{We are assuming here for simplicity that all
5-branes have their transverse coordinates located at the
orbifold fixed point at the origin. This leads to the 
configuration with maximal gauge symmetry and is also
explicitly invariant under T-dualities.}. Thus the above configuration 
is equivalent, by a T-duality transformation with respect to all compact 
coordinates, to the case of one 3-brane and three 7-brane sectors. One can 
switch between the two configurations by replacing respectively 
$(9,5_1,5_2,5_3)$ with $(3,7_1,7_2,7_3)$. One can of course consider models 
or configurations in which some of the p-brane types are absent, like
e.g., a model with only 9(3)-branes or a model with only 9(3)-branes
and one set of $5_i(7_i)$-branes etc.
In this case one has just to delete from the formulae below 
the fields corresponding to the non existing sectors.
Let us then concentrate on the general case with one 9-brane sector
and three $5_i$-brane sectors. Then there will be generically
gauge groups $G_9$, $G_{5_i}$ corresponding to those four sectors.
These groups will not be in general semisimple. Now, there will be
the following types of charged matter fields:

\noindent 1) $C_i^9$ , {\it i}=1,2,3  

\noindent 
Here $i$ labels the three complex compact dimensions.
They come from open strings starting and ending on 9-branes and 
have only quantum numbers with respect to
the gauge group $G_9$. They transform in general 
as a reducible representation of $G_9$ .
The three of them ({\it i}=1,2,3) transform differently with respect to
$G_9$ (except for the $Z_3$ orientifold). As emphasized in \cite{afiv}
these
fields behave quite similarly to untwisted sectors of perturbative
heterotic orbifolds.

\noindent 2)  $C_j^{5_i}$, $i$,$j$=1,2,3 

\noindent Analogous to the previous
ones but
with $9$-branes replaced by $5_i$-branes, they come from open strings 
starting and ending on the same $5_i$-brane and are only 
charged with respect to the $G_{5_i}$ group. There are three of them
($j$=1,2,3)  for each given $5_i$-brane.

\noindent 3) $C^{95_i}$, $i$=1,2,3 

\noindent They come from strings
starting (ending) 
on a 9-brane and ending (starting) on a $5_i$-brane. They have
gauge quantum numbers under both $G_9$ and $G_{5_i}$ (typically
transform as bifundamental representations). Since the open string 
has Neumann boundary conditions at the 9-brane but Dirichlet boundary
conditions
along the compact dimensions transverse to the $5_i$ world-volume,
they behave in some way like the $Z_2$ twisted sectors of 
even order perturbative heterotic orbifolds.

\noindent 4) $C^{5_i5_j}$, $i$,$j$=1,2,3, $i\not= j$ 

\noindent They come from open strings
starting
in one type of $5_i$ brane and ending on a different type ($i\not= j$)
of $5_j$-brane. They are analogous to the previous ones but
with $9$-branes replaced by $5_j$-branes. They have gauge quantum numbers
under both $G_{5_i}$ and $G_{5_j}$. 

In a given particular model not all these types of charged fields
might be present. Thus for example, in a model with one sector of
9-branes and one sector (e.g. $5_1$) of 5-branes the last type of
fields are absent. Let us remark again that the same general classes of
charged fields exist for the T-dual configuration with one set of
3-branes and 3 sets of $7_i$-branes doing the obvious replacements.  

In order to present the general form of the $D=4$ $N=1$
effective Lagrangian is convenient to consider first  a model 
with  9-branes only. In $D=10$ the effective Type I field theory Lagrangian
will be similar to the one of the $SO(32)$ heterotic, except for
the different dilaton dependence of gauge couplings. Then the effective
Lagrangian in
$D=4$  for a compact orientifold with only 9-branes 
will be completely analogous to the one of untwisted particles 
in a heterotic orbifold. Thus the gauge kinetic function $f_9$ and
K\"ahler
potential $K$ will have the general form \cite{rev} :
\beqa
f_9 & = & S \ ,
\label{nueva}
\\
K & = & -\log  (S + S^*)   \ - \
          \sum_{i=1}^3 \log \left(T_i+T_i^* - |\cni|^2  \right) \ ,
\label{solo9}
\eeqa
where the complex chiral fields $S$, $T_i$ with $i$=1,2,3 are now given by
\cite{afiv}
\beqa
S & = & {{2 R_1^2R_2^2R_3^2}\over {\lambda_I (\alpha ' )^3 } }  
  + i\theta \ ,
\label{nueva2}
\\
T_i & = & {{2R_i^2}\over
{\lambda _I\alpha '}} + i\eta _i  
\ ,
\label{sttt}
\eeqa
where $\theta $ and $\eta _i$ are untwisted RR closed string states
and $\lambda _I$ is the 10-dimensional dilaton.
Notice that formulae (\ref{nueva}) and (\ref{nueva2}) with $Re f_9=4\pi/g_9^2$  
reproduce the second  term in
(\ref{d04}). 
Now, invariances under T-dualities allow us to reconstruct the complete
$f$-functions and K\"ahler potential for the more general situation with 
three sets of 5-branes $5_i$, $i$=1,2,3.
In particular let us consider T-duality transformations 
$D_{ij}$ with respect to two
complex planes $X_i,X_j$
simultaneously. There are three such transformations:
\beq
D_{ij}\  :\ \   (R_i,R_j) \ \longrightarrow (\alpha '/R_i, \alpha '/R_j )
  \ , \ i\not= j \ ,
\label{dualidades}
\eeq
for ($i$,$j$)=(1,2), (1,3), (2,3). Under these 3 transformations 
$ReS$ and $ReT_i$ are permuted  into each other
(see eqs.(\ref{rescaleo}, \ref{nueva2}, \ref{sttt})). In particular
one obtains that $D_{ij}$ act on $(ReS,ReT_1,ReT_2,ReT_3)$ as matrices:
\beq
D_{23}\  = \bmat{cc}\sigma _1& 0 \\ 0  & \sigma _1  \emat
\quad ; \quad
D_{13}  = \bmat{cc}0& \id_{2}\\ \id_{2} & 0 \emat \quad ;\quad D_{12}=D_{23}
D_{13} \ ,
\label{tresduales}
\eeq
where $\sigma_1$ is the Pauli matrix. Simultaneously the type of
D-branes transform under these three symmetries accordingly i.e.,
the $D_{ij}$ permute the four type of branes $(9,5_1,5_2,5_3)$ with the
same matrices as above. Thus e.g., under a $D_{23}$ transformation one has:
\beqa
S &\leftrightarrow  &T_1 \ ,\nonumber\\
T_{2,3}&\leftrightarrow  &T_{3,2} \ ,\nonumber \\ 
  {\rm D}9  &\leftrightarrow  & {\rm D}5_1 \ ,\nonumber \\
  {\rm D}5_{2,3} &\leftrightarrow & {\rm D}5_{3,2} \ .
\label{duales23}
\eeqa
Consider for example a situation with only 9-branes and one set of 
$5_1$-branes. We know that a configuration in which all transverse 
dimensions of the $5_1$-branes sit at the fixed point at the origin 
should be explicitly self-dual under these transformations. Thus making 
eqs.(\ref{nueva}, \ref{solo9}, \ref{nueva2}, \ref{sttt}) invariant will require 
gauge kinetic functions $f_9,f_{5_1}$ and K\"ahler potential given by
\cite{afiv} :
\beqa
f_9\ &  = & \ S \quad\quad , \quad\quad f_{5_1}\ =\ T_1 \ ,
\nonumber \\[0.2ex]
K &=& -\log(S+S^* - |C_1^{5_1}|^2)   -  \log(T_1+T_1^* - |C_1^9|^2 )
\nonumber \\[0.2ex]
   & - & \log(T_2+T_2^*-|C_2^9|^2 -|C_3^{5_1}|^2 )  -
\log(T_3+T_3^*-|C_3^9|^2 -|C_2^{5_1}|^2 )
\nonumber\\[0.2ex]
  &+ & {{|C^{95_1}|^2}\over {(T_2+T_2^*)^{1/2}(T_3+T_3^*)^{1/2}}} \ , 
\label{kali}
\eeqa
which is explicitly invariant.
As remarked in ref. \cite{afiv}, 
 the form of the metric of the
charged $C^{95_1}$ fields is suggested by  T-duality invariance and the
fact that the $(95_1)$ sector of these theories behaves as a sort
of $Z_2$ twisted sector with twist under the dualized directions
$X_2,X_3$. Indeed open strings with Neumann conditions in one end and
Dirichlet ones on the other get a $Z_2$ twist for those dimensions
\cite{gp,bl} .

In the more general situation in which we have 9-branes and all three
types of $5_i$-branes, invariance under the $D_{ij}$ dualities gives the
results
for gauge kinetic functions:
\beqa
f_9\ &  = & \ S \quad\quad , \quad\quad f_{5_i}\ =\ T_i \ .
\label{kineticf}
\eeqa
So differently from the weakly coupled heterotic string theories now the
(tree-level)
gauge kinetic functions are not independent of the moduli sector\footnote{In
the weakly coupled heterotic case  a dependence from the moduli sector can
arise only
at loop level and so it is very small. This is not however the case of
the
strongly coupled heterotic from $M$--theory \cite{Horava}
where both contributions are of the same order \cite{Funcion,Nilles}, 
$f\sim S+T$.}.
For the K\"ahler potential one obtains:
\beqa
K & = & -\log \left(S + S^* - \sum_{i=1}^3 |\ccii|^2 \right) \ - \
         \sum_{i=1}^3 \log \left(T_i+T_i^* - |\cni|^2 - 
\sum_{j,k=1}^3 d_{ijk} |\ccjk|^2 \right) \nn \\
  &+& \frac{1}{2}\sum_{i,j,k=1}^3d_{ijk} 
{|\ccjck|^2 \over {(S+S^*)^{1/2} (T_i+T_i^*)^{1/2}}} + 
  \frac{1}{2} \sum_{i,j,k=1}^3 d_{ijk} 
{|\cnci|^2 \over {(T_{j}+T_{j}^*)^{1/2}
                                      (T_{k}+T_{k}^*)^{1/2}}}\ , \nn
\\
\label{kahlershort}
\eeqa
with $d_{ijk}=1$ if $i \neq j \neq k \neq i$, otherwise 
$d_{ijk}=0$. 
Expanding  
the K\"ahler potential in the lowest order in the matter fields we obtain the  
following expression:
\beqa
K & = & - \log(S + S^*) - \sum_{i=1}^3 \log(T_i + T_i^*) \ + \
   \sum_{i=1}^3 {|\cni|^2 \over {(T_i + T_i^*)}} \nn \\ 
&+&
        \sum_{i=1}^3 {|\ccii|^2 \over {(S + S^*)}} \ + \
{{|\ccjk|^2} \over {(T_i + T_i^*)}} \nn \\ 
&+& 
     \frac{1}{2}  \sum_{i,j,k=1}^3  
d_{ijk} {|\ccjck|^2 \over {(S+S^*)^{1/2} (T_i+T_i^*)^{1/2}}} 
 + \frac{1}{2}\sum_{i,j,k=1}^3  
d_{ijk} {|\cnci|^2 \over {(T_{j}+T_{j}^*)^{1/2}
                                       (T_{k} + T_{k}^*)^{1/2}}} \ . \nn
\\
\label{kahlerex}
\eeqa
This is explicitly invariant under the full set of $D_{ij}$ dualities.
The first piece in (\ref{kahlerex}) is the usual term corresponding to the
complex dilaton field $S$ that is present for any compactification. The second
term is also in general present in the orbifold compactification of
weakly coupled heterotic
strings, while all the other pieces are characteristic of the Type I
case
(as mentioned above, the matter field contribution from the 9-brane
sector,
third piece, is analogous to the one of the untwisted matter in a
heterotic orbifold). The forth piece is similar to the one appearing,
due to higher order corrections, in the strongly coupled heterotic 
from $M$--theory \cite{ovrut}. 

One can also write down the form of the superpotential
for this more general situation. It can be derived knowing that
these renormalizable interactions come from joining and splitting
of open strings ending on the different types of D-branes \cite{bl}.
One gets in the most general configuration:
%
%
\beqa
W & = & g_9 \left(\cna \cnb \cnc \ + \ \ccacb \cccca \ccbcc \ + \ 
\sum_{i=1}^3 \cni \cnci \cnci \right) \ + \  \sum_{i,j,k=1}^3 g_{5_i} 
\left(\ccia \ccib \ccic \right. \nn \\
 & & \ + \ \left. \ccii \cnci \cnci \ + \ d_{ijk} \ccij \ccick \ccick 
\ + \ \frac{1}{2} d_{ijk}  \ccjck \cncj \cnck \right) \ .
\label{superpot}
\eeqa
%
The Yukawa coupling  constants are given by:
\beq
g^2_9 \ =\ { {4\pi }\over {Re S} }\ \ ,\ \ 
g^2_{5_i}\ =\  { {4\pi }\over {Re T_i} } \ .
\label{yukaw}
\eeq

Two general remarks are in order:
\begin{itemize}
\item
If some type of D-brane is absent in the orientifold under study,
one has to delete the corresponding chiral fields in the above expressions.

\item 
Under a T-duality transformation with respect to the three 
compact dimensions the 9-branes transform into 3-branes and the
$5_i$-branes into $7_i$-branes. Still the effective Lagrangian 
will be identical to the one above with the obvious replacements 
$9\rightarrow 3$, $5_i\rightarrow 7_i$ everywhere. Notice however that
now one has differently defined $S$ and $T_i$ fields with:
\beq
ReS\ = \ {2\over {\lambda _I} } \ \ ;\ \ 
ReT_i\ =\ { {2R_j^2R_k^2}\over {\lambda _I \alpha '^2} }\ , \ 
i\not = j\not = k\not = i
\label{otrosttt}
\eeq
in agreement with eq.(\ref{gaugci}).
\end{itemize}
%
%
\section{Soft SUSY-breaking terms from dilaton/moduli spurions}
%
%
\subsection{General p-brane configurations}
We will now consider the point of view of refs. \cite{IL,KL,BIM} in studying 
soft SUSY-breaking terms in low-energy field theories
from the perturbative heterotic string. In those references
it was argued that the complex dilaton $S$ and moduli
fields $T_i$ in $N=1$, $D=4$ orbifolds vacua are natural candidates 
for the SUSY-breaking hidden sector. If that is the case
SUSY breaking could be analyzed in terms of
the vevs of the dilaton and moduli auxiliary fields 
%
\begin{eqnarray}
F^S &=& \sqrt{3} C m_{3/2}(S+S^*)\sin\theta
        e^{-i\gamma _S} \ , \nn \\
F^i &=& \sqrt{3} C m_{3/2}(T_i + T_i^*)\cos\theta
        \Theta_{i} e^{-i\gamma_i}\ ,
\label{auxili}
\eea
where we are using the parametrization introduced in \cite{BIM,BIMS} in
order to know what fields, either $S$ or $T$, play the predominant role
in the process of SUSY breaking. 
The angle $\theta$ and the $\Theta_i$ with $\sum_i |\Theta_i|^2=1$,
just parametrize the direction of the goldstino in the $S$, $T_i$ field
space, $m_{3/2}$ is the gravitino mass, 
$C^2=1+\frac{V_0}{3m_{3/2}^2}$ with $V_0$ the tree-level vacuum energy
density, and $\gamma_S$, $\gamma_i$ are the phases of 
$F^S$ and $F^i$.
%
%
%

In this section we would like to carry out a similar
analysis for the effective $N=1$, $D=4$ field theories
from Type IIB orientifolds described above.
We would also like to emphasize that the above assumption of 
the dilaton/moduli fields transmitting SUSY breaking to
the observable charged particles is even more compelling
in the context of Type I $D=4$, $N=1$ string models. 
Indeed, if, as described in the section 5, SUSY is
broken in some hidden sector of e.g., 3-branes, only
closed string fields can transmit this symmetry breaking to the
observable D-brane sector since only
closed strings can move in the bulk.  In Type IIB orientifolds the
only massless chiral fields in the closed string sector
are the dilaton $S$ and the moduli $T_i$. In addition there
are twisted moduli singlet fields $M_a$ who couple only 
locally to the branes close to the singularities. Thus the untwisted 
moduli fields $S$, $T_i$ are singled out to transmit any SUSY-breaking 
effect, if present.

Applying the standard (tree-level) soft term formulae \cite{SW,BIM22}
for the effective Lagrangian described in the previous section 
(\ref{kineticf}, \ref{kahlerex}, \ref{superpot}), we can compute the soft
terms straightforwardly. Since the bilinear parameter, $B$, depends on the
specific mechanism which could generate the associated $\mu$ term, let us
concentrate on 
the other soft parameters. After normalizing the observable fields, the
gaugino masses are given by
%
\bea
M_9 & = & \sqrt{3} C m_{3/2} \sin \theta e^{-i \gamma _S} \ , \nn \\
M_{5_i} & = & \sqrt{3} C m_{3/2} \cos \theta \Theta_i e^{-i \gamma _i} \ .
\label{gaugino1}
\eea
%
%
The scalar masses are given by
\bea
m^2_{\cni} & = & m^2_{\ccjk} = 
m_{3/2}^2 + V_0  - 3 C^2 m_{3/2}^2 \cos^2 \theta \Theta_{i}^2 = 
m_{3/2}^2 + V_0 - |M_{5_i}|^2 
\ , \nn \\
m^2_{\ccii} & = &  m_{3/2}^2 + V_0  - 3 C^2 m_{3/2}^2 \sin^2 \theta = 
m_{3/2}^2 + V_0 - |M_{9}|^2 
\ , \nn \\
m^2_{\cnci} & = &  m_{3/2}^2 + V_0  - \frac{3}{2} C^2 m_{3/2}^2 
\cos^2 \theta \left(1 - \Theta_{i}^2 \right) =
m_{3/2}^2 + V_0 -\frac{1}{2}\left( |M_{5_j}|^2 + |M_{5_k}|^2 \right)
\ , \nn \\
m^2_{\ccicj} & = & m_{3/2}^2 + V_0  - \frac{3}{2} C^2 m_{3/2}^2
\left(\sin^2\theta + \cos^2\theta \Theta_{k}^2 \right)  = 
m_{3/2}^2 + V_0 -\frac{1}{2}\left( |M_{9}|^2 + |M_{5_k}|^2 \right)
\ . \nn \\
\label{scalar1}
\eea
%
%
Finally the results for the trilinear parameters are
%
\bea
A_{C^9_1 C^9_2 C^9_3} & = & A_{ C_i^{9} C^{95_i} C^{95_i}} =
   -\sqrt{3} C m_{3/2}\sin\theta e^{-i \gamma _S}= - M_9 
\ , \nn \\
A_{C_1^{5_i}C_2^{5_i}C_3^{5_i}} & = & A_{C_i^{5_i} C^{95_i}C^{95_i}} =
   A_{\ccij \ccick \ccick} = -\sqrt{3}Cm_{3/2} \cos\theta \Theta _i
   e^{-i \gamma _i}  = - M_{5_i} 
\ , \nn \\
A_{\ccacb \cccca \ccbcc} & = & \frac{\sqrt 3}{2}Cm_{3/2}
   (\sin\theta e^{-i \gamma_S} - \cos\theta \sum_i\Theta _i e^{-i \gamma_i})
   = {1 \over 2} (M_9 - \sum_{i=1}^3 M_{5_i}) 
\ , \nn \\
A_{\ccicj \cnci \cncj} & = &  \frac{\sqrt 3}{2}Cm_{3/2}
   \left[\cos\theta \left(\Theta_k e^{-i \gamma_k} - \Theta_i e^{-i \gamma_i} 
  - \Theta _j e^{-i \gamma_j}\right) - \sin\theta e^{-i \gamma_S}\right] \nn \\
 & = & {1 \over 2} ( M_{5_k} - M_{5_i} - M_{5_j} - M_9) \ ,
\label{trilin1}
\eea
with $i,j,k = 1,2,3$ and $i \neq j \neq k \neq i$ in the above equations.
Notice that the last equalities in the formulae written in
(\ref{scalar1}) and (\ref{trilin1}) make sense {\it only if} 
the corresponding gauginos are in fact present in the spectrum (i.e., if the
corresponding D-branes are there). If that is not the case we see that we 
cannot write the corresponding scalar mass or trilinear parameter in terms 
of just gaugino masses, we get an explicit dependence on $\theta$ and 
$\Theta _i$.
%

Finally, for three particles coupled through a Yukawa, using
(\ref{scalar1}) with $V_0=0$ ($C=1$), one finds that, in general for
{\it any} choice of the goldstino direction, the following relevant
sum-rules are fulfilled:
\bea
m^2_{\cna}+m^2_{\cnb}+m^2_{\cnc} & = & m^2_{\cni}+ m^2_{\cnci} +  m^2_{\cnci} 
    = |M_9|^2 
\ , \nn \\
m^2_{\ccia}+m^2_{\ccib}+m^2_{\ccic} & = & m^2_{\ccii}+ m^2_{\cnci} + 
    m^2_{\cnci} = m^2_{\ccij}+ m^2_{\ccick} + m^2_{\ccick} =  
    |M_{5_i}|^2 
\ , \nn \\
m^2_{\ccjck}+m^2_{\cncj}+m^2_{\cnck} &=& |M_9|^2+|M_{5_j}|^2+|M_{5_k}|^2
    - {3\over2} m^2_{3/2} 
\ , \nn \\
m^2_{\ccacb}+m^2_{\cccca}+m^2_{\ccbcc} &=& 
    \sum_{i=1}^3 |M_{5_i}|^2 - {3\over2} m^2_{3/2} \ .
\label{sumrules}
\eea
with $i,j,k = 1,2,3$ and $i \neq j \neq k \neq i$. 
Notice that in general the gravitino mass appears
in some of these sum-rules. In the case in which all four
types of D-branes are present one can eliminate that 
explicit dependence by using (see (\ref{gaugino1}))
\be 
|M_9|^2 + \sum_{i=1}^3 |M_{5_i}|^2 = 3 m^2_{3/2} \ .
\label{gaugra}
\ee
%
%
%

With the above information we can analyze now the structure of soft
terms available for each possible D-brane configuration as well as to
compare
it with that of  perturbative heterotic vacua \cite{BIM22}
or  $M$--theory models \cite{Nilles,Mua,Ovrut2}.

\subsection{ Configurations with 9-branes and one set of 5-branes}

Let us consider first  the situation when 9-branes and 
one set of $5_i$-branes, say $5_1$-brane,
are simultaneously present in the model. This is in
fact
a representative example which will allow us to extract a number of
generic qualitative properties of soft terms in orientifold models.
Besides, this is a particularly interesting case because if we look at it
from the perspective of the S-dual heterotic string it corresponds to
the presence of both perturbative heterotic fields (dual to the
9-brane sector) and non-perturbative heterotic fields (dual to
the 5-brane sector). Thus this case may in principle show us 
some features of non-perturbative heterotic vacua. In this case the
soft terms, assuming $V_0=0$ (C=1) and $\gamma_S=\gamma_i=0$ (mod $\pi$)
given the current experimental limits, are:
\bea
M_9^2 & = & 3 m^2_{3/2} \sin^2 \theta \ , \nn \\
M_{5_1}^2 & = & 3 m^2_{3/2} \cos^2 \theta \Theta_1^2 \ , 
\label{mia1}
\eea
\bea
m^2_{C_1^9} & = & m_{3/2}^2 (1 - 3 \cos^2 \theta \Theta_1^2) \ , \nn \\
m^2_{C_2^9} & = & m^2_{C_3^{5_1}} = m_{3/2}^2 (1- 3 \cos^2 \theta
\Theta_2^2) \nn \ , \\
m^2_{C_3^9} & = &  m^2_{C_2^{5_1}} = m_{3/2}^2 (1 - 3 \cos^2 \theta
\Theta_3^2) \nn \ ,
\\
m^2_{C_1^{5_1}} & = & m_{3/2}^2 (1 - 3 \sin^2 \theta) \nn \ , \\
m^2_{C^{9{5_1}}} & = & m_{3/2}^2 \left[1-\frac{3}{2} \cos^2 \theta
                    \left(1 - \Theta_{1}^2 \right) \right]\ , 
\label{mia2}
\eea 
\bea
A_{C^9_1 C^9_2 C^9_3} & = & A_{C^9_1 C^{9{5_1}} C^{9{5_1}}} = - M_9 \ , \nn \\
A_{C_1^{5_1} C_2^{5_1} C_3^{5_1}} & = &   A_{C_1^{5_1} C^{9{5_1}} C^{9{5_1}}}
= - M_{5_1} 
\ .
\label{mia3}
\eea
The following sum-rules are satisfied:
\bea
m^2_{C_1^9} + m^2_{C_2^9} + m^2_{C_3^9} &=& m^2_{C_1^9} + 
m^2_{C^{9{5_1}}} + m^2_{C^{9{5_1}}} = M_9^2 \ , \nn \\
m^2_{C_1^{5_1}}+m^2_{C_2^{5_1}}  + m^2_{C_3^{5_1}} &=& 
m^2_{C_1^{5_1}}+m^2_{C^{9{5_1}}}  + m^2_{C^{9{5_1}}} = M_{5_1}^2 \ .
\label{mia4}
\eea
Notice that the structure of these soft terms is qualitatively different 
from that of the (untwisted sector of the) perturbative heterotic
string\footnote{It is also different from that of the strongly coupled
heterotic from $M$--theory \cite{Mua} where both $\sin\theta$ and 
$\cos\theta$ contribute together to all soft parameters. In this sense, 
Type I soft terms are an ``average'' of weakly coupled and strongly 
coupled heterotic soft terms.} \cite{BIMS}:
\bea
M^2 & = & 3 m^2_{3/2} \sin^2 \theta \ , 
\label{bobo1}\\
m^2_{C_i} & = & m_{3/2}^2 (1 - 3 \cos^2 \theta \Theta_i^2 )\ , 
\label{bobo2}\\
A_{C_1 C_2 C_3} & = & - M \ ,
\label{bobo3}
\eea
with the sum-rule
\be
m^2_{C_1} +m^2_{C_2}  + m^2_{C_3}  = M^2 \ .
\label{sum9}
\ee
Within the Type I string framework here described, these results
would correspond
to the limiting case when there are no SM fields in the $5_1$-brane
sector, see (\ref{mia1}, \ref{mia2}, \ref{mia3}, \ref{mia4})
(or equivalently to the D-brane configuration with only 9-branes
present).
In that case one could extract a number of generic qualitative
properties
of soft terms with regard to three important issues: the universality
of soft terms, the existence or not of negative squared mass for some
matter fields, and the relative sizes of gaugino versus scalar masses.
One finds \cite{BIM,BIMS} :

\noindent 1) Unlike gaugino masses (\ref{bobo1}),  scalar masses
(\ref{bobo2}) are generically non universal. Universality may only be
obtained in two cases: First, in the dilaton-dominated SUSY breaking
($\sin^2 \theta = 1$) which implies $m^2_{C_i}= m_{3/2}^2$. 
Second, in the overall modulus case
($\Theta_1=\Theta_2=\Theta_3=\frac{1}{\sqrt 3}$) which
implies $m^2_{C_i}= m_{3/2}^2 \sin^2 \theta$. This situation is shown
in Fig.\ref{fig2a} with solid lines (only $C_i^9$ fields are present).
\begin{figure}[t]
\begin{center}
\epsfig{file=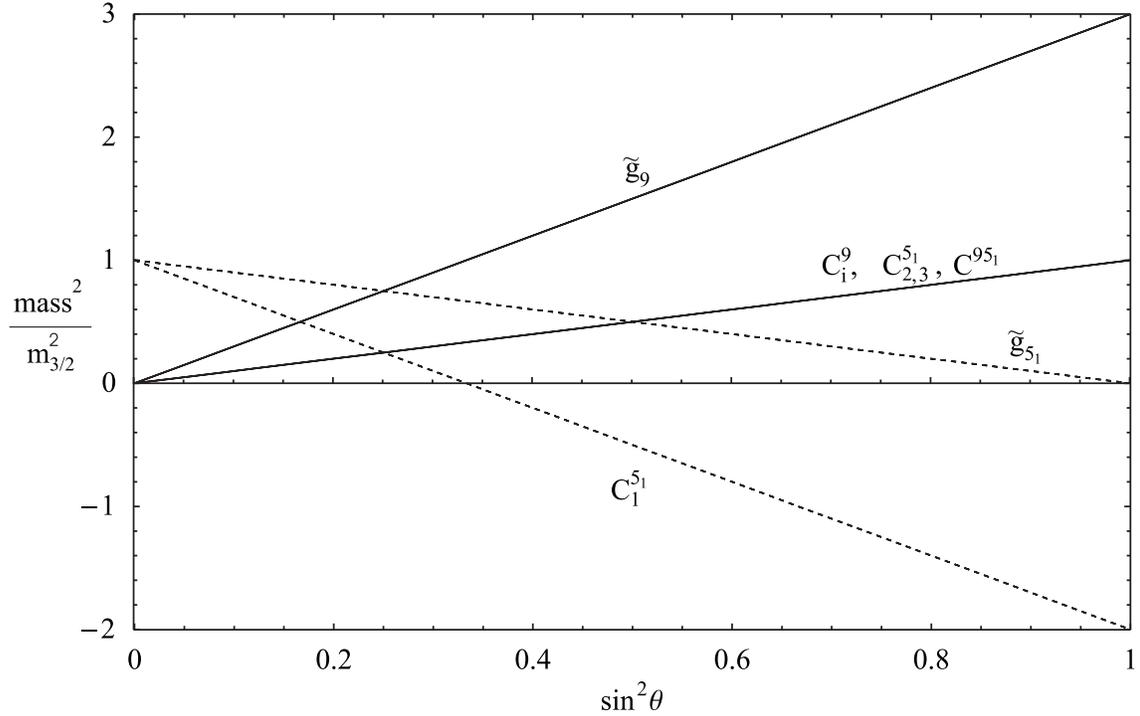,height=9.5cm} 
\end{center}
\caption{\it \small
Scalar (C) and gaugino ($\tilde{g}$) squared masses in unit 
of $m^2_{3/2}$ versus $\sin^2 \theta$ for S/overall modulus ($\Theta_i = 
1/\sqrt{3}$) SUSY breaking when $9$-branes and one set of $5_1$-branes are 
present. The solid lines (with only the scalar fields $C_i^9$) correspond 
to the situation where only $9$-brane sectors are present.}
\label{fig2a}
\end{figure}
\begin{figure}[t]
\begin{center}
\epsfig{file=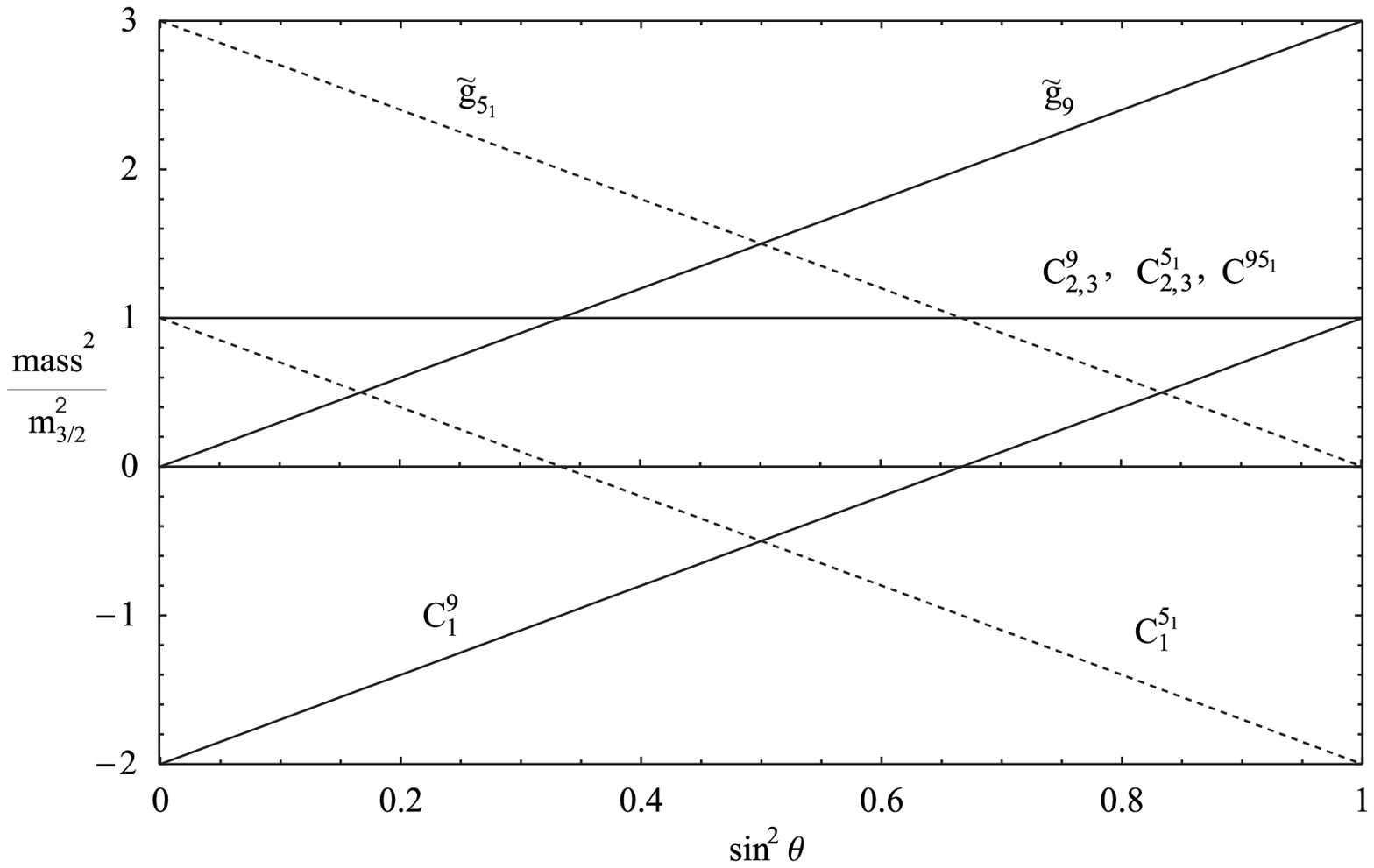,height=9.5cm} 
\end{center}
\caption{\it \small 
Scalar (C) and gaugino ($\tilde{g}$) squared masses in unit 
of $m^2_{3/2}$ versus $\sin^2 \theta$ for S/$T_1$ ($\Theta_1 =1$) 
SUSY breaking when $9$-branes and one set of $5_1$-branes are present. 
The solid lines (with only the scalar fields $C_i^9$) correspond to the 
situation where only $9$-brane sectors are present.}
\label{fig2b}
\end{figure}

\noindent 2) One immediately observes from (\ref{bobo2})
that the presence of tachyons can be avoided for {\it any}
choice of the goldstino direction by imposing the condition
$\cos^2 \theta \leq 1/3$. This is shown in Fig.\ref{fig2b} with solid lines
(only $C_i^9$ fields are present) for the interesting example
$\Theta_1=1, \Theta_{2,3}=0$, i.e. $S/T_1$ SUSY breaking.

\noindent 3) The above sum-rule (\ref{sum9}) implies that scalars
heavier than gauginos can exist at the cost of having some of the other 
three scalars with negative squared mass\footnote{Nevertheless,
exceptions can arise in several situations and besides having a
tachyonic sector is not necessarily a potential problem as discussed in 
\cite{BIMS}.}.

We would like now to study to what extent these conclusions change in
the more general case given by (\ref{mia1}-- \ref{mia4}) in which
there are charged fields both in 9-branes and 5-branes.
We insist that this case corresponds to the S-dual of 
non-perturbative heterotic vacua with a perturbative gauge 
group (dual to that coming from 9-branes)
and a non-perturbative group (dual to that coming from 5-branes).

\noindent 1) {\it Universality of soft masses} 

\noindent Now, not only scalar masses (\ref{mia2}) but also 
gaugino masses (\ref{mia1}), due to the different values of $f_\alpha$ in 
the different sectors (\ref{kineticf}), are
generically non universal. Neither in the dilaton limit,  
$\sin^2 \theta = 1$, nor in the overall modulus limit,
$\Theta_{1,2,3}=\frac{1}{\sqrt 3}$, universality can be obtained. This 
can be seen in Fig.\ref{fig2a} where the dependence on 
$\sin^2 \theta$ of the soft squared masses in orientifold models when 
9-branes and one set of $5_1$-branes are present is showed. For
example, for  $\sin^2 \theta = 1$, gauginos in the 9-brane sector,
$\tilde {g}_9$, have 
$|M_9|=\sqrt 3 m_{3/2}$ whereas gauginos in the $5_1$-brane sector,
$\tilde {g}_{5_1}$, 
have\footnote{This limit is also interesting in the
context of SUSY breaking by gaugino condensation.
In the perturbative heterotic string $f=S$ and therefore the minimum
of the scalar potential is obtained for 
$F^S\propto \sin\theta\approx 0$, with the result that (tree-level)
gaugino masses are very small 
$M\propto F^S\approx 0$ \cite{Beatriz}.
However, in type I strings, 
if we assume that the
condensation takes place e.g. in a hidden sector of $5_1$-branes,
then $f_{5_1}=T_1$, i.e. $F^1$ has the same role as $F^S$ above:
$F^1\propto \cos\theta\approx 0$. As a result although we get 
$M_{5_1}\approx 0$, in the visible sector of 9-branes $M_9\propto F^S\propto 
\sin\theta \neq 0$ implying in general large gaugino masses $\approx
m_{3/2}$. This can also be obtained in  
the case of the strongly coupled 
heterotic string from M--theory  \cite{Nilles} since there 
$f\sim S+T$.}
$M_{5_1}=0$. Besides, all scalars have a squared mass equal to $m^2_{3/2}$
except $C_1^{5_1}$ which has a negative squared mass $-2 m^2_{3/2}$.

One interesting  point to emphasize is that there is a value  $\sin^2\theta
=1/4$
which is T-duality self-invariant in this overall modulus situation. 
Indeed there one has $\frac {F^S}{S+S^*}=\frac{F^i}{T+T^*}$. 
At that point $|M_9|=|M_{5_1}|=\frac{\sqrt 3}{2} m_{3/2}$
and all scalars of all sectors have $m=\frac{1}{2}m_{3/2}$.

\noindent 2) {\it Tachyons}

\noindent The presence of tachyons {\it cannot} be avoided for any
choice of the goldstino direction. For example, in the overall modulus
limit we see in Fig.\ref{fig2a} that there are no tachyons for 
$\sin^2 \theta \leq 1/3$. For larger values $C_1^{5_1}$ becomes
tachyonic.
However, in Fig.\ref{fig2b} we show the direction in which only $S$ and $T_1$
contribute
to SUSY breaking where either 
$C_1^9$ or $C_1^{5_1}$ (or both) become tachyonic.
Let us remark however that this is not necessarily a problem if there
are no SM particles in those sectors. 

Finally, it is worth noticing that this direction is interesting because 
the soft terms explicitly show (see Fig.\ref{fig2b}) the invariance under 
T-duality which exchanges $S\leftrightarrow T_1$. Thus here dilaton 
dominance and modulus dominance are dual to each other.

\noindent 3) {\it Gaugino versus scalar masses}

\noindent Now due to the possibility of having gauginos in different
D-brane sectors, 9-brane and $5_1$-brane, the sum-rules 
(\ref{mia4}) imply that scalars heavier than gauginos can be possible
without other scalars becoming tachyonic. We can have 
$m_{C_i^9} > |M_{5_1}|$ or 
$m_{C_i^{5_1}} > |M_9|$. For example, in Fig.\ref{fig2a} for  
$\sin^2 \theta = 0$ (modulus-dominated limit)
we obtain $m_{C_1^{5_1}}=m_{3/2} > M_9=0$ with
$m_{C_{2,3}^{5_1}}= m_{C_i^9} =m_{C^{9{5_1}}}=0$ 
and $M_{5_1}=m_{3/2}$.\\

Comparing these conclusions for two types of branes (9, $5_1$) with
those summarized above for the perturbative heterotic string
(which let us remark again are equivalent to a Type I string with only
9-branes present) one certainly finds plenty of differences.
Configurations with three and four types of branes, e.g. (9, $5_1$, $5_2$) 
and (9, $5_1$, $5_2$, $5_3$), give rise to similar results. The only
difference arises in the configuration with only one set of
$5_i$-branes, say $5_1$-brane. There the existence of universal gaugino 
masses $M_{5_1}$ implies that the sum-rule
$m^2_{C_1^{5_1}} + m^2_{C_2^{5_1}}  + m^2_{C_3^{5_1}} = M_{5_1}^2$ 
(see (\ref{mia4})) is fulfilled, without some scalars becoming tachyonic,
only if scalars are lighter than gauginos.
%
%
\section{Anomalous $U(1)$'s, twisted moduli and precocious gauge 
coupling unification}
%
%
In the class of Type IIB, $D=4$, orientifold models under discussion
there are pseudoanomalous $U(1)$ interactions in the spectrum.
Such pseudoanomalous $U(1)$'s are also present in heterotic 
perturbative vacua although they have different properties than in the
orientifold case, as we now discuss. 

Let us first recall how are things in the perturbative heterotic case.
In that case $D=4$, $N=1$ vacua have at most one anomalous $U(1)_X$.
The anomaly is canceled by the 4-dimensional version
\cite{dsw,ckm,sin} of the Green-Schwarz mechanism \cite{gs}. 
In the latter an important role is played 
by the imaginary part of the complex heterotic dilaton $S$.
Under an anomalous $U(1)_X$ gauge transformation with parameter
$\theta (x)$ it gets shifted by $c\theta (x)$, c being a constant.
Since the Lagrangian contains the couplings 
$Im S \sum_a k_a F_a\wedge F_a$, where the sum runs over all gauge groups in
the model, a shift in $ImS$ can in principle cancel mixed
$U(1)_X$-gauge anomalies. However, for this to be possible the 
mixed anomalies have to be in the same ratios as the coefficients
$k_a$ (Kac-Moody levels) of the gauge factors. This is 
an important constraint which has lead to interesting 
phenomenological applications \cite{sin,ibaross}. 
In addition, a Fayet-Iliopoulos (FI)  term
for the scalar potential associated to the anomalous $U(1)_X$ is also
generated at one-loop in string theory \cite{dsw,fihet}. 
This FI term is proportional to $\Tr Q_X g/(192\pi ^2)$.

In the case of toroidal $D=4$, $N=1$ Type IIB orientifolds it
has been found \cite{iru} that the situation is quite different. 
The main characteristics of anomalous $U(1)$'s in this case 
are as follows:

{\bf i)  } 
There are {\it multiple anomalous $U(1)$'s}  and their mixed anomalies with
the rest of the gauge groups are {\it not universal}, i.e., they are
not in the ratios of the corresponding coupling constants.

{\bf ii)} There is a generalized Green-Schwarz mechanism at work which
cancels all mixed anomalies \cite{iru} 
. In this mechanism the complex dilaton $S$
does not play any role and so happens also with the untwisted moduli
fields $T_i$. It is the twisted  fields  $M_a$ associated to the
fixed points of the underlying orbifold which participate
in the cancellation. Furthermore, the required terms appear at the tree
(disk) level and not at the one-loop level as in the heterotic case.
 Specifically, there are two kind of couplings 
which conspire to get the cancellations. The first of them has the
general form:
\beq
\sum _k d^l_k B_k \wedge F_{U(1)_l} 
\label{shift}
\eeq
where $k$ runs over the different twisted sectors of the 
underlying orbifold (see ref.\cite{iru} for details) and 
$B^k$ is the two-form which is the dual to the imaginary
part of the twisted fields $M_a$. Here {\it l} labels the different
anomalous $U(1)$'s and $d^l_k$ are model-dependent constant 
coefficients\footnote{ Specifically one has
$d^l_k=Tr(\gamma_k \lambda ^l)$ where $\lambda ^l$ is the 
Chan-Paton matrix associated to the anomalous $U(1)$ and
$\gamma _k$ is the action of the orbifold twist on 
Chan-Paton factors.}. This is a mixing term which connects
the anomalous $U(1)$'s with the twisted RR fields associated to the
fixed points. The second coupling which participates in the
anomaly cancellation is $M_a$-dependent piece which appears 
at the tree-level for the gauge kinetic functions $f_{\alpha }$
corresponding to the different gauge groups $G_{\alpha }$.
Thus e.g. for a gauge group coming from p-branes one has
in general:
\beq
f_{\alpha }\ = \ f_p \ + \ \sum _k s^{\alpha }_k M_k
\label{correc}
\eeq
where $f_p$ are the gauge kinetic functions described in section 6 
(see e.g. eq.(\ref{kineticf}) ) and $s^{\alpha }_k$ are model dependent 
coefficients\footnote{In these class of orientifolds one has 
$s^{\alpha}_k=Tr(\gamma_k^{-1}(\lambda_{\alpha })^2)$ where
$\lambda _{\alpha }$ is the Chan-Paton matrix corresponding to the
$G_{\alpha }$ group.}. The $M_k$ are twisted closed string 
chiral fields corresponding to the $k$-th twisted sector.
The imaginary part of these $M_k$ fields are the duals of the
$B_k$ two-form fields in eq.(\ref{shift}).

The generalized Green-Schwarz mechanism works here as follows. 
An anomalous $U(1)_l$ gauge boson mixes with the imaginary part
of a twisted $M_k$ field (or, equivalently, its dual $B_k$ form)
as in eq.(\ref{shift}). Then the latter propagates and couples to two
$G_{\alpha }$ gauge bosons through the second piece in eq.(\ref{correc}).
This is quite analogous to the mechanism in the perturbative
heterotic case, the main difference  being that there are several
fields (the twisted $M_a$ fields) participating in the mechanism.
This allows for different  mixed anomalies for the different gauge factors.
Unfortunately in the present case the details are model dependent
since the number of $M_k$ fields and the $d^l_k$, $s^{\alpha }_k$
coefficients are indeed model dependent. For some simple examples see
ref.\cite{iru}.

{\bf iii)}
The presence of the mixing term (\ref{shift}) and supersymmetry require
also the presence of a FI term for each of the anomalous $U(1)$'s
\cite{dm,iru}:
\beq
D_l\ \sum _k d^l_k ReM_k
\label{FI}
\eeq
where $D_l$ is the auxiliary field for each of the anomalous $U(1)_l$ and
$ReM_k$ are related to the blowing-up modes (twisted moduli)
which repair the orbifold singularities. 
Thus the value of the FI terms  is perturbatively undetermined 
and  vanish in the orbifold limit. It has been shown \cite{poppitz} that 
the one-loop corrections to these FI-terms vanish.

{\bf iv) }
The mass of the anomalous $U(1)$'s depend on the form of the
Kahler potential for the twisted fields $M_k$. Assuming
a Kahler potential of the form $K_M=(M_k+M_k^*)^2$ it is easy to check
\cite{poppitz}  
that all anomalous $U(1)$'s get a mass of order the string scale.
If the Kahler potential turns out to have a more complicated
form, the mass of the given $U(1)$ would depend on the size
of the corresponding FI term and hence could be both heavy or light.
Notice that in the first case the anomalous $U(1)$'s would be massive
even if  the FI terms vanish. Thus in this case the anomalous $U(1)$
would remain like perturbatively exact global symmetries\footnote{Notice 
that this is {\it not} the case in general perturbative heterotic models 
because in that case the FI term cannot be put to zero and hence 
further gauge symmetry breaking is induced which generically also
break the would-be remnant global $U(1)$ symmetry.}. This could be 
important in order to insure proton stability in particular string models.

The consequences of the above described FI terms for the dynamics
of the models are again very model-dependent since they depend on the
twisted sectors of the theories. Nevertheless 
we would like  to point out some 
other possible phenomenological consequences
of the above results. 

{\it  a) Gauge coupling  non-universality and
precocious unification }

The twisted moduli $M_k$ dependent piece in the gauge kinetic functions
can have important implications. Notice in particular that this piece is
different  
for different gauge groups, even if all group factor come from the 
same set of p-branes\footnote{Notice that for a twisted
moduli field $M_k$ to appear in the gauge kinetic function there
must be some overlap between the p-brane world-volume and the
corresponding fixed point. Thus, e.g, in the case of 3-branes
they must be sitting close to the singularity.}. Away from the orbifold
limit (i.e., $\langle ReM_k\rangle \not= 0$ ) 
the gauge couplings will be different and hence
{\it there is no universality of gauge couplings}.
Of course, we already saw in section 3 that if gauge groups 
come from different types of p-branes there is already 
no universality of gauge couplings. What we are saying now is that
even if all gauge groups come from the same collection of p-branes, 
the gauge couplings could be different as long as we are  close to
(but not on top of) the orbifold singularity.

On the other hand this non-universality could be {\it interesting in
obtaining gauge coupling unification in  low
string scale models}. For example, in the scheme of ref.\cite{BIQ} 
with a string scale $M_I\approx \sqrt{M_WM_{Planck}}$ one has to achieve
gauge coupling unification also at that scale which is of order
$10^{11}$ GeV. The addition of extra chiral fields in the
massless spectrum to achieve precocious gauge coupling unification
is a possibility \cite{BIQ}. Here we would like to point out another
option which naturally appears in the present context.
Consider the renormalization group running of gauge couplings 
$g_{\alpha}$ from the weak scale to the string scale $M_I$:
\beq
{{4 \pi}\over {g_{\alpha}(M_W) }}\ = \ {Re f_{\alpha}}
\ + \ { {  b_{\alpha } }\over {2\pi}} \log{ {M_I}\over {M_W} }
\label{running}
\eeq
where $f_{\alpha }$ is the gauge kinetic function in eq.(\ref{correc}).
We know that with the particle content of the MSSM coupling
unification works nicely for a unification scale $M_X=2\times
10^{16}$ GeV. Now consider a simplified situation with only one 
relevant $M_k$ twisted field. Eq.(\ref{correc}) would have the form
\beq
f_{\alpha } = S \ + \ s_{\alpha }M \ .
\label{correca}
\eeq
Thus if we had a model with:
\beq
 s_{\alpha } =  \gamma b_{\alpha }  \ ; \quad  
\langle  Re M \rangle \ = \ {1 \over \gamma} {1 \over {2\pi}} \log(M_X/M_I)
\label{precoz}
\eeq
we would nicely get gauge coupling unification.

It turns out that some odd order compact orientifolds ($Z_3$, $Z_7$) 
have the required interesting property that $s_{\alpha} \propto b_{\alpha}$, 
where in this case $M$ is the sum over all the twisted moduli fields. 
Let us describe  the $D=4$, $N=1$, $Z_3$ orientifold \cite{ang} 
as an example. This model contains only 9-branes\footnote{Of course, 
T-dual models can be constructed which have only either 3,5 or 7-branes 
instead.} and has gauge group $U(12)\times SO(8)$ and charged
chiral fields transforming as $3(12,8_v)_1$$+3({\overline {66}},1)_{-2}$,
where the subindex shows the $U(1)$ charges.
The beta functions of the $SU(12)$ and $SO(8)$ groups are
respectively $-9$ and $+18$. On the other hand the coefficients 
$s_{SU(12)}$ and $s_{SO(8)}$ in this model are given by 
$\cos2\pi V_{\alpha }$ (see eq.(3.4) in ref.\cite{iru}), where 
$V_{SU(12)}=2/3$ and $V_{SO(8)}=0$. Thus we have $s_{SU(12)}=-1/2$ 
and $s_{SO(8)}=+1$, indeed proportional to the beta functions. Something 
analogous does also happen in the $Z_7$ orientifold and also if we add 
Wilson lines to the models.

These are just specific examples and it is not clear to
what extent this property could be more general.
Nevertheless the  orientifold examples show that
indeed it is not unreasonable the above scheme for gauge coupling 
unification for models with low string scales.  
In principle it should also be applicable to models with $M_I=1$ TeV 
where the problem of precocious gauge unification is more difficult to 
obtain by other means. Of course, a realistic model with the required 
properties remains to be built. 

{\it  b) Corrections to SUSY-breaking soft terms }

The results obtained in the previous section for gaugino
masses did not take into account the extra piece 
(\ref{correc}) which can appear in the context 
of  orientifold models. Of course, those formula remain valid
as long as the twisted $M_k$ fields do not participate 
in the process of SUSY breaking, i.e., $F_{M_k}=0$.
One important point to realize is that these twisted
fields live only close to each singularity, they do not
leave in the bulk of the six compact dimensions. Thus
they do not participate in the transmission of SUSY breaking from 
one (hidden) p-brane sector to another (visible) p-brane
sector.  Thus, unlike what one would have expected, having $F_{M_k}\not=0$ 
in a hidden p-brane sector does not in general gives rise to 
non-universalities in gaugino masses since the $M_k$ fields which couple 
to differently positioned p-branes are different fields\footnote{Thus 
e.g., if we have two sets of 3-branes one corresponding to a visible 
sector and the other to a hidden sector, sitting at
different singularities, each sector will couple to
the twisted fields $M_k$ corresponding to each 
corresponding singularity.}. However, although universality
is not affected, the soft mass terms are modified with respect to the
results in the previous section because now in general
the goldstino field will have a $M_k$ component also.

The role of these $M_k$ twisted fields in soft terms is
somewhat similar to that considered in chapter 8 of ref.\cite{BIM}.
There it was studied the effect of fields contributing to
SUSY breaking but not coupling to the visible world.
One then has to introduce an extra goldstino angle.
This fact was already pointed out in ref.\cite{BIQ}.

Let us end  this section by emphasizing that the
effects discussed here are present in 
$D=4$, $N=1$ Type IIB orientifold models and need not
be present perhaps in more general Type I constructions. 
In particular, it turns out that in the case of 
{\it smooth} Calabi-Yau compactification of the $D=10$ Type I string, 
there is again a single anomalous $U(1)$ whose anomaly is canceled
by a Green-Schwarz  mechanism analogous to the one in
perturbative heterotic vacua. In particular, it is the 
exchange of the complex dilaton field $S$ which cancels the anomaly.
On the other hand it seems likely that, whenever we have sets of
e.g., 3-branes sitting close to some sort of 
(not necessarily an orbifold) singularity
similar effects will be present.
%
%
\section{Final comments and outlook}
%
%
In the present paper we have studied a number of
properties of Type IIB, $D=4$, $N=1$ orientifolds
which can be phenomenologically relevant. We 
consider these vacua as interesting laboratories
to explore the phenomenological consequences of
a D-brane view of the unification of the
standard model and gravity. More general vacua
which may involve other Type IIA or Type IIB
D-brane configurations which go beyond toroidal
orbifolds may perhaps be necessary for a final
realistic description of the real world, but we
hope that some of the generic features found in toroidal 
orientifolds could still be present in more general situations.

One can conceive more general models which will still be describable 
in terms of different sets of  Dp-branes (one of them
containing the SM) sitting either at singularities
of the compact space or in the bulk. Chirality of
the SM seems to suggest that the Dp-brane set 
describing the SM should be sitting close to
some singularity in compact space, Dp-branes
in the bulk give rise generically to 
vector-like spectra. Those singularities 
need not be simple orbifold singularities.

A generic vacuum may also contain non-supersymmetric Dp-brane sectors 
coming e.g. from Dp-branes wrapping on non-supersymmetric cycles of 
the compact six-dimensional space \cite{sennew}. If that is the case, 
lowering the string scale well below the Planck scale may be
{\it not just possible but mandatory}. Indeed, if we have a vacuum with
a $N=1$ p-brane sector containing the SM and other $N=0$ sectors away 
from it and with a string scale close to $M_{Planck}$, SUSY-breaking terms 
will appear in the visible $N=1$ SM sector which are not going to be 
sufficiently suppressed. In this case, setting
$M_I \leq \sqrt{M_WM_{Planck}}$ will be in general required.

The possibility of decreasing the string scale changes
also the perspective concerning the generation of
the $M_W/M_{Planck}$ hierarchy. It was shown in ref.\cite{BIQ} 
and we have discussed in more detail in section 5
that in Type I $D=4$ vacua there appear factors of
the form $\alpha ^2(M_c/M_I)^6$ for this hierarchy.
Thus for $M_c<<M_I$ one can reproduce the required suppression.
The choice of value for $M_I$ which minimizes
fine-tuning (in the sense of requiring not too small
ratio,  $M_c/M_I\approx \alpha \approx 10^{-2}$) is the geometric scale
$\sqrt{M_WM_{Planck}}$. This assumes a scheme in which 
SUSY breaking is transmitted from a $N=0$ sector to the
visible $N=1$ SM sector by fields living in the bulk. 

In a Type I scheme in which SUSY breaking is transmitted from
a hidden $N=0$ brane sector to a $N=1$ brane sector 
containing the SM, the natural candidates for this transmission
are the dilaton/moduli fields of the model. Specifically,
in the context of Type IIB toroidal $D=4$, $N=1$ 
orientifolds other closed string chiral fields like the
twisted moduli live closed to the orbifold singularities,
not in the bulk. Thus it seems sensible to parametrize
SUSY breaking in the visible $N=1$ sector in terms
of dilaton/moduli spurions as we have described in section 
7 in the spirit of refs.\cite{BIM,BIMS}. 

Lowering the string scale has its shortcomings. One of them
is the question of gauge coupling unification. In a scheme
with the string scale of order $M_X=2\times 10^{16}$ GeV,
gauge coupling unification is nicely achieved.
In section 3 we have studied in detail how that
happens for different Dp-brane configurations. On the other
hand the hierarchy $M_W/M_{Planck}$ needs to have its origin in
a specifically assumed hierarchy-generating mechanism like
gaugino condensation. In schemes with a low string scale
(lower than $M_X$) one expects the running of SM couplings to stop at 
the string scale where they are not yet unified. In section 8 we have
suggested a possibility to understand this lack of unification
at the string scale. If the SM p-branes sit close to  a singularity
(which anyway seems to be required in order to get chirality)
there are tree-level contributions to the gauge couplings which
are generically different for each gauge group. If these
contributions are proportional to the beta-function of each 
group they split the couplings in the appropriate way to
understand lack of unification. It turns out that there
exist specific Type IIB $D=4$, $N=1$ orientifolds in which 
indeed these contributions to gauge couplings are 
proportional to the beta functions. This at least shows that
this mechanism is a new possibility to understand the
question of gauge coupling unification in low string scale
models. If this mechanism is indeed at work, it would mean
that the branes at which the SM lives sat close to 
singularity in the compact space.

Another potential problem for low string scale models is nucleon
stability. The lower the string scale, the higher the
dimension of possible operators one has to worry about.
Anyway, dimension=4,5 are a problem in all schemes, including
those with $M_I=M_X$. In all cases there must be extra 
continuous or discrete (gauge symmetries) suppressing 
appropriately proton decay. A number of possibilities
both for discrete and continuous gauge symmetries 
ensuring proton stability were studied in ref.\cite{IR} .
Moreover, Type II string vacua seem to have generically 
abundant $U(1)$ symmetries, most of them anomalous 
which could play an important role in ensuring proton stability.

In summary, recent developments in string dualities
have changed in many ways our views concerning how
to embed the observed physics inside string theory.
Surely there will be more surprises waiting for us.

\bigskip

\bigskip

\noindent {\bf Acknowledgements}

We thank C. Burgess, A. Donini, F. Quevedo , R. Rabadan, A. Uranga and 
G. Violero for discussions. 

S. Rigolin acknowledges the European Union for financial 
support through contract ERBFMBICT972474.


\begin{thebibliography}{99}
%
\bibitem{rev}
For reviews and references, see:\\
F. Quevedo, hep-ph/9707434; hep-th/9603074;\\
K. Dienes, hep-th/9602045;\\
J. Lykken, hep-th/9607144;\\
G. Aldazabal, hep-th/9507162;\\
L. E. Ib\'a\~nez, hep-th/9505098;\\
Z. Kakushadze and S.-T. Tye, hep-th/9512155.
%
\bibitem{polrev}
For a review, see: J. Polchinski, hep-th/9611050.
%
\bibitem{witten}
E. Witten, \NPB {471} {96} {135}, hep-th/9602070.
%
\bibitem{lykken}
J.D. Lykken, Phys. Rev. D54 (1996) 3693, hep-th/9603133.
%
\bibitem{untev}
N. Arkani-Hamed, S. Dimopoulos and G. Dvali, hep-ph/9803315.
%
\bibitem{anton}
I. Antoniadis, N. Arkani-Hamed, S. Dimopoulos and G. Dvali,
hep-ph/9804398; \\
I. Antoniadis, S. Dimopoulos, A. Pomarol and M. Quiros, hep-ph/9810410.
%
\bibitem{bajogut}
K. Dienes, E. Dudas and T. Gherghetta, hep-ph/9803466;
hep-ph/9806292; hep-ph/9807522.
%
\bibitem{gr}
D. Ghilencea and G.G. Ross, hep-ph/9809217.
%
\bibitem{sundrum}
R. Sundrum,
hep-ph/9805471; hep-ph/9807348.
%
\bibitem{shiutye}
G. Shiu and S.H. Tye, hep-th/9805157.
%
\bibitem{bachas}
C. Bachas, hep-ph/9807415.
%
\bibitem{kakutye}
Z. Kakushadze and S.H. Tye, hep-th/9809147.
%
\bibitem{benakli}
K. Benakli, hep-ph/9809582.
%
\bibitem{bendav}
K. Benakli and S. Davidson, hep-ph/9810280;\\
D.H. Lyth, hep-ph/9810320.
%
\bibitem{bursto}
N. Arkani-Hamed, S. Dimopoulos and G. Dvali, hep-ph/9807344;\\
P. Argyres, S. Dimopoulos and J. March-Russell, hep-th/9808138;\\
K. R. Dienes, E. Dudas, T. Gherghetta and A. Riotto, hep-ph/9809406;\\
N. Arkani-Hamed, S. Dimopoulos and J. March-Russell, hep-th/9809124;\\
L. Randall and R. Sundrum, hep-th/9810155.
%
\bibitem{BIQ}
C. Burgess, L.E. Ib\'a\~nez and F. Quevedo, hep-ph/9810535.
%
\bibitem{kakutev}
Z. Kakushadze, hep-th/9811193.
%
\bibitem{bursta}
M. Maggiore and A. Riotto, hep-th/9811089;\\
G. Giudice, R. Rattazzi and J.D. Wells, hep-ph/9811291;\\
S. Nussinov and R. Shrock, hep-ph/9811323;\\
T. Han, J.D. Lykken and R.J. Zhang, hep-ph/9811350;\\
E.A. Mirabelli, M. Perelstein and M.E. Peskin, hep-ph/9811337;\\
N. Arkani-Hamed and S.  Dimopoulos, hep-ph/9811353;\\
J. Hewett, hep-ph/9811356;\\
Z. Berezhiani and G. Dvali, hep-ph/9811378;\\
K.R. Dienes, E. Dudas and A. Gherghetta, hep-ph/9811428;\\
N. Arkani-Hamed, S. Dimopoulos and J. March-Russell,
hep-ph/9811448;\\
P. Mathews, S. Raychaudhuri and K. Sridhar, hep-ph/9811501.
%
\bibitem{dhvw}
L.~Dixon, J.A.~Harvey, C.~Vafa and E.~Witten, \NPB{261} {85} {678}; B274 
1986 285.
%
\bibitem{orbi}
L.E.~Ib\'a\~nez, J.~Mas, H.P.~Nilles and F.~Quevedo,
\NPB{301}{88}{157};\\
A.~Font, L.E.~Ib\'a\~nez, F.~Quevedo and A.~Sierra,
\NPB{331}{90}{421}.
%
\bibitem{bl}
M.~Berkooz and R.~G.~Leigh, \NPB{483} {97} {187}, hep-th/9605049.
%
\bibitem{ang}
C.~Angelantonj, M.~Bianchi, G.~Pradisi, A.~Sagnotti and Ya.S.~Stanev,
\PLB{385} {96} {96}, hep-th/9606169.
%
\bibitem{3kakus}
Z.~Kakushadze, \NPB {512} {98} 221, hep-th/9704059;\\
Z.~Kakushadze and G.~Shiu, \PRD{56} {97} {3686}, hep-th/9705163;\\
Z.~Kakushadze and G.~Shiu, Nucl. Phys. B520 1998 75, hep-th/9706051.
%
\bibitem{zwart}
G.~Zwart, Nucl. Phys. B526 (1998) 378, hep-th/9708040.
%
\bibitem{odri}
D.~O'Driscoll, hep-th/9801114.
%
\bibitem{fin}
L.E.~Ib\'a\~nez, hep-th/9802103.
%
\bibitem{afiv}
G. Aldazabal, A. Font, L.E. Ib\'{a}\~{n}ez and 
G. Violero, FTUAM-98/4, hep-th/9804026.
%
\bibitem{2kakus}
Z. Kakushadze, hep-th/9804110; hep-th/9806044.
%
\bibitem{lpt}
J. Lykken, E. Poppitz and S. Trivedi, hep-th/9806080.
%
\bibitem{BIM22} For a recent review, see:
A. Brignole, L.E. Ib\'{a}\~{n}ez and C. Mu\~noz,
in the book `Perspectives on Supersymmetry', Ed. G. Kane, (World
Scientific, 1998) p. 125; hep-ph/9707209.
%
\bibitem{IL} L.E. Ib\'{a}\~{n}ez and D. L\"ust,
Nucl. Phys. B382 (1992) 305.
%
\bibitem{KL} V.S. Kaplunovsky and J. Louis,
Phys. Lett. B306 (1993) 269.
%
\bibitem{BIM} A. Brignole, L.E. Ib\'{a}\~{n}ez and C. Mu\~noz,
Nucl. Phys. B422 (1994) 125 [Erratum: B436 (1995) 747].
%
\bibitem{BIMS} A. Brignole, L.E. Ib\'{a}\~{n}ez, C. Mu\~noz and C.
Scheich, Z. Phys. C74 (1997) 157.
%
\bibitem{iru}
L. E. Ib\'a\~nez, R. Rabad\'an and A. Uranga, hep-th/9808139 .
%
\bibitem{ibanez}
L.E. Ib\'a\~nez, hep-ph/9804236.
%
\bibitem{sagnotti}
A.~Sagnotti, in Cargese 87, `Strings  on Orbifolds',
Ed. G. Mack et al. (Pergamon Press, 1988) p. 521.
%
\bibitem{hor}
P.~Horava, \NPB{327} {89} {461}; \PLB{231} {89} {251};\\
J.~Dai, R.~Leigh and J.~Polchinski, Mod.Phys.Lett. A4 (1989) 2073;\\
R.~Leigh, Mod.Phys.Lett. A4 (1989) 2767.
%
%
\bibitem{bs} G.~Pradisi and A.~Sagnotti, \PLB{216} {89} {59};\\
M.~Bianchi and A.~Sagnotti, \PLB{247} {90} {517}.
%
\bibitem{gp}
E.~Gimon and J.~Polchinski,
Phys.Rev. D54 (1996) 1667, hep-th/9601038.
%
\bibitem{dp1}
A.~Dabholkar and J.~Park, \NPB{472} {96} {207}, hep-th/9602030.
%
\bibitem{gj}
E.~Gimon and C.~Johnson,
\NPB{477}{96}{715}, hep-th/9604129.
%
\bibitem{dab}
For an introduction to orientifolds see, e.g.: 
A. Dabholkar, hep-th/9804208.
%
\bibitem{witpol}
J. Polchinski and E. Witten, \NPB{460} {96} {525} , hep-th/9510169.
%
\bibitem{tv}
T. Taylor and G. Veneziano, \PLB {212} {88} {147}.
%
\bibitem{Horava} P. Horava and E. Witten, Nucl. Phys. B460
(1996) 506, hep-th/9510209; B475 (1996) 94, hep-th/9603142. 
%
\bibitem{Funcion} T. Banks and M. Dine, Nucl. Phys. B479
(1996) 173, hep-th/9605136;
\\
K. Choi, Phys. Rev. D56
(1997) 6588, hep-th/9706171;
\\
H.P. Nilles and S. Stieberger, Nucl. Phys. B499 (1997) 3, hep-th/9702110.
%
\bibitem{Nilles}
H.P. Nilles, M. Olechowski and M. Yamaguchi, 
Phys. Lett. B415
(1997) 24, hep-th/9707143.
%
\bibitem{ovrut}
A. Lukas, B.A.Ovrut and D. Waldram, Nucl. Phys. B532 (1998) 43, hep-th/9710208.
%
\bibitem{SW} S.K. Soni and H.A. Weldon, Phys. Lett. B126 (1983) 215.
%
%
\bibitem{Mua} K. Choi, H.B. Kim and C. Mu\~noz, Phys. Rev. D57 
(1998) 7521, hep-th/9711158.
%
\bibitem{Ovrut2}
A. Lukas, B.A. Ovrut and D. Waldram, Phys. Rev. D57 (1998) 7529, 
hep-th/971119.
%
\bibitem{Beatriz} B. de Carlos, J.A. Casas and C. Mu\~noz, Phys. Lett.
B299 (1993) 234.
%
\bibitem{dsw}
M. Dine, N. Seiberg and E. Witten, \NPB{289} {87} {585}.
%
\bibitem{ckm}
J. Casas, E. Katehou and C. Mu\~noz, \NPB{317} {89} {171}.
%
\bibitem{sin}
L.E. Ib\'a\~nez, \PLB{303} {93} {55}.
%
\bibitem{gs}
M. Green and J. Schwarz, \PLB{149} {84} {117}.
%
\bibitem{ibaross}
L.E. Ib\'a\~nez and G.G. Ross, \PLB{332} {94} {100};
for a review and references, see: P. Ramond, hep-ph/9604251.
%
\bibitem{fihet}
J. Atick, L. Dixon and A. Sen, \NPB{292} {87} {109};\\
M. Dine, I. Ichinoise and N. Seiberg, \NPB{293} {87} {253}.
%
\bibitem{dm}
M. Douglas and G. Moore, hep-th/9603167.
%
\bibitem{poppitz}
E. Poppitz, hep-th/9810010.
%
\bibitem{sennew}
A. Sen, hep-th/9812031.
%
\bibitem{IR}
L.E. Ib\'a\~nez and G.G. Ross,
\NPB{368} {92} {3}.
%
\end{thebibliography}
\end{document}